\documentclass[12pt,preprint]{aastex}

\shorttitle{Mid-IR interferometry of massive stars}
\shortauthors{Rajagopal et al.}

\begin{document}


\title{Mid-Infrared interferometry of dust around massive evolved stars \\
    }


\author{Jayadev Rajagopal\altaffilmark{1}, Jean-Luc Menut\altaffilmark{2}, 
D. Wallace\altaffilmark{3}, W.C. Danchi\altaffilmark{3},
O. Chesneau\altaffilmark{4}, B. Lopez\altaffilmark{2},
J.D. Monnier\altaffilmark{5}, M. Ireland\altaffilmark{6}, P.G. Tuthill\altaffilmark{7}}
\email{jrajagopal@ctio.noao.edu}


\altaffiltext{1}{Michelson Fellow, University of Maryland, College Park MD 20742, \\
Present address: CTIO, Casilla 603, La Serena, Chile}
\altaffiltext{3}{NASA Goddard Space Flight Center, Greenbelt, MD 20771, USA}
\altaffiltext{2}{Observatoire de la C\^ote d'Azur, D\'epartment G\'emini, UMR 6203, Nice, France}
\altaffiltext{4}{Observatoire de la C\^ote d'Azur, D\'epartment G\'emini, Avenue Copernic, Grasse, 
France}
\altaffiltext{5}{Department of Astronomy, University of Michigan, 501
East University Avenue, Ann Arbor, MI 4809, USA}
\altaffiltext{7}{School of Physics, Sydney University, NSW, 2006, Australia}
\altaffiltext{6}{Planetary Science MC 150-21, California Institute of Technology, Pasadena CA 91125}


\begin{abstract}
We report long-baseline interferometric measurements of circumstellar 
dust around massive 
evolved stars with the MIDI instrument on the Very Large Telescope 
Interferometer and provide spectrally dispersed visibilities in the 8-13 \micron\ 
wavelength band. We also present diffraction-limited observations at
10.7 \micron\ on the Keck Telescope with baselines up to 8.7 m which
explore larger scale structure. We have resolved the dust shells around 
the late type WC stars 
WR 106 and WR 95, and the enigmatic NaSt1 (formerly WR 122), suspected to   
have recently evolved from a Luminous 
Blue Variable (LBV) stage. For AG Car, 
the protoypical 
LBV in our sample, we marginally resolve structure close to the star, 
distinct from the well-studied detached nebula.  
The dust shells  around the two WC stars show fairly 
constant
size in the 8-13 \micron\ MIDI band, with gaussian half-widths of $\sim$ 25 to 
40 mas, and the  Keck observations 
reveal an  additional extended structure around WR 106. 
The visibility profiles for NaSt1 obtained from two MIDI baselines 
indicate a compact source embedded in an extended structure. The compact
dust we detect around NaSt1 and AG Car favors recent or ongoing dust formation.

Using the measured
visibilities, we 
build spherically symmetric radiative transfer models of the WC
dust shells which enable detailed comparison with existing SED-based models. 
Our results indicate that the inner radii of the shells are
within a few tens of AU from the stars. 
In addition, our models favor grain size distributions with large
($\sim$ 1 \micron) dust grains.
This proximity of the inner dust to the hot central star emphasises
the difficulty faced by current theories in forming dust in the hostile
environment around WR stars. 
Although we detect no direct evidence for binarity for these objects, dust 
production in a 
colliding-wind interface in a binary system is a feasible
mechanism in WR systems under these conditions. 

\end{abstract}





\section{Introduction}
The most massive stars ($\gtrsim 40 $M$_\sun$) will likely pass through a 
Luminous Blue Variable (LBV) and Wolf-Rayet (WR) phase before ending their lives
in a supernova explosion. LBVs and some WRs are known to show infrared 
excess associated with circumstellar dust. The LBV phase is short-lived and
unstable with irregular mass ejections of $10^{-5}$ to $10^{-4} $M$_\sun$ for
each outburst \citep{hum94}, resulting in circumstellar nebulae 
and
associated dust. The dust in this phase is consistent 
with grains of differing compositions ranging over metallic, silicate and 
carbon-rich \citep{voo99,cla03}.  The WR stage, characterised by massive 
winds (of the order $\dot{\rm M} = 10^{-5}\rm M_\sun {\rm yr}^{-1}$) is divided into two 
main phases \citep{smi68}, 
the WN phase with
strong spectral lines of N and CNO-cycle products, and the later WC phase 
with strong lines of C and other He-burning products.
Dust formation 
is associated with the WC phase, consistent with the over-abundance of C, and is
notably absent in the WN phase \citep{vdh01b}, although recent {\it Spitzer} observations
\citep{bar06} may be the first indications to the contrary. Long term and detailed 
infrared photometry of WC stars \citep[hereafter WvdHT]{will87},  has
led to considerable progress in understanding the nature of the dust and its 
formation. The favored mechanism for dust-formation in this stage
is through wind-wind collision in a WR-OB binary system \citep{uso91}. For
the episodic dust producers WR 140 and WR 137, photometric and spectrosocopic 
data \citep{will90,will01} had
already indicated binarity which has been confirmed by long-baseline 
interferometric observations for WR 140 \citep{mon04a}. 
High resolution images in the near-IR using aperture-
masking techniques with the 
Keck telescope provide direct
evidence for binarity in the case of the constantly dusty, late-type WC stars 
WR 98a, WR 104 \citep{mon99,tut99}, and more recently, the ``cocoon stars'' in the 
Quintuplet cluster \citep{tut06}. 
\Citet{mon07} present a strong case for binarity for all the dusty WRs in the Keck 
aperture-masking sample.  However, whether all 
dusty late type WCs are binaries is an open question and dust formation in 
single WR stars has not been
ruled out \citep{che00}.       

In this paper, we present and interpret interferometric mid infrared 
observations of the      
late type WC stars \object{WR 95} and \object{WR 106}, the suspected 
LBV-WR transition
star \object{NaSt1}, and the LBV \object{AG Car}. The main goals of this 
campaign were
to resolve the sizes of the dust shells and characterize the physical 
distribution of dust. All the objects were observed with the MIDI instrument
at the Very large Telescope Interferometer (VLTI) 
in the 8-13 \micron\ wavelength range, typically with a baseline of $\sim$ 47 m. 
NaSt1 and WR 106 were 
also observed at 10.7 \micron\ with a multi-aperture interferometric technique 
using individual
segments of the Keck-I telescope mirror to get complementary 
data from baselines up to the 10 m diameter of the Keck mirror. 
In Sections 2 and 3, we describe the observations, data reduction and 
present the results. 
Section 4 details the dust models we have constructed and the
comparison with existing models based on the spectral energy distribution.
Section 5 summarizes these results and presents the conclusions.

\section{Observations and Data Reduction}
\subsection{MIDI}
Most targets were observed over four nights in 2004 July, except for AG Car 
which was allocated one night in 2004 April (Table 1). NaSt1 was also 
observed on two nights in 2004 September. 
All observations were with the $\sim$ 47 m long UT2-UT3 baseline, except for the
2004 September observations of NaSt1 which used the $\sim$ 89 m UT2-UT4 baseline.
The MIDI instrument 
\citep{lei03} on the VLTI combines the light from a given 
pair of 8.2 m telescopes (UTs)
and produces spectrally dispersed fringes in the 8-13 \micron\ band. A detailed
description of the observing technique and data reduction can be found in
\citet{lei04}. Here we limit ourselves to a brief summary.

The observing sequence starts with acquiring the target on each
individual telescope with the 8.7 \micron\ filter (1.4 \micron\ wide) and attaining 
maximal overlap of the two images, while chopping the secondary at
$\sim$ 2 Hz over 10\arcsec. 
The images are 
combined to
produce the two complementary interferograms and 
dispersed through a prism with a resolution of $\lambda/\delta\lambda \sim$ 30. 
The delay lines are scanned to find and track the fringes 
and fringe visibilities are recorded while stepping the delay through $\sim$
10 wavelengths ($\lambda = $10 \micron) in steps of 2 \micron\ with a piezo-driven mirror.
The scan rate is tuned to
the timescale of instrumental and atmospheric fluctuations of delay and
a typical measurement records
$\sim$ 200 scans in about 5 minutes. After the fringe measurement, a photometric
measurement is carried out by blocking the light from one and then the other
telescope, while chopping the secondary mirror. No chopping is done during
the fringe-tracking scans. The pointing is maintained by the Coude guiding with
only tip-tilt correction; higher order AO was yet to be integrated into the system.
A sensitivity better than $\sim$ 1 Jy is achieved for this
mode of fringe measurement. Given that these observations were during the
first regular semester after MIDI commisioning, the accuracy that we achieve
for the visibility measurements is of the order of 10\% and is
mostly set by photometry, guiding (overlap of the two images) and 
fringe tracking errors which impose limits on the accuracy of calibration. 
We calculate the errors on the visibility as the standard deviation
over the multiple calibrations for each measurement and conservatively quote 
the higher of this number or 10\% of the visibility amplitude. 
The above is true for the {\it absolute} visibility amplitude errors; the
relative errors are less and the slope of the visibility in the 8 to
13 \micron\ band is accurate to $\sim$3\%.

Standard calibrator stars of known diameter from the MIDI catalogue\footnote{  
\url[HREF]{http://www.eso.org/projects/vlti/instru/midi/midi\_calib.txt}} 
were observed
immediately before and after the target (Table 1). Given the relatively large systematic
errors inherent in the visibility determination, we have also used calibrators from 
other programs within a 
span of 2-5 hours (depending on atmospheric conditions).    

The visibility reduction method generally follows the now well-established algorithms
for optical/IR interferometry (e.g., \citet{for97} for fiber stellar interferometers: the
basic principles apply for non-fiber systems like MIDI) 
and was carried out using custom software
provided by the VLTI consortium written in the {\it IDL} language. The spectra
are extracted with a typical binning of 4 pixels, with each pixel being 
$\sim$ 0.05 \micron\ wide. 
The two complementary
outputs of the combiner are differenced to remove the incoherent flux including
overall variations in 
the contribution from the two telescopes. Next, the fringe power is estimated
from the Fourier transform and the visibility amplitude\footnote{The usual estimator
used in optical/infrared interferometry is the square of the visibility 
amplitude (V$^2$) since
this quantity can be corrected for detector noise bias. However the MIDI 
measurement error in the mode we used (``High-Sens'', with no simultaneous photometry) is dominated by calibration error, 
not detector noise.}, normalized by the photometric fluxes, is calculated.
The visibility is then calibrated using the instrument visibility measured 
on the calibrator stars. 

\subsection{Keck}
We carried out observations at 10.7 \micron\ with the Keck-I telescope 
on two stars in our sample, WR 106 and NaSt1, in 2005 May (Table 2). These observations
were carried out in a ``segment-tilting'' mode \citep{mon04b,wei06} to achieve close to diffraction
limited resolution. 
Imaging at 10 \micron\ with the full pupil has, in
practice, failed to reliably yield diffraction limited images. This
is in part attributable to the ``seeing spike" problem, where the
visibilities at shorter baselines are difficult to calibrate in the
presence of changes in the seeing between the target and calibrator.
The results shown here demonstrate the utility of this new method in
alleviating some of these problems. In this technique, 
custom software is used to  
reconfigure the segmented primary mirror so as to allow selected groups of
segments to focus on the detector (the Long-Wavelength Spectrograph camera). 
Each group (four in our case) of
6 segments is chosen so as to form a non-redundant sparse-aperture 
array. The 1.8 m hexagonal segments are individually
phased up so that each group focuses on a separate spot on the detector. 
Thus in a given frame, 24 out of 36
segments are used in these four ``Fizeau arrays'' (the rest of the segments are
pointed away from the detector). This arrangement measures 15 baselines
ranging from about 2 m to 9 m and 20 closure phases. Each frame is of
the order of 10 ms and we co-add 9 frames, giving an
effective exposure of 90 ms. Chopping is done at a 5 Hz rate. Calibrators 
are observed before and
after the targets and each target is visited at least twice during
a night. The analysis of
the data closely follows the aperture-masking analysis described in 
detail in numerous publications \citep[e.g.,][]{tut00}.
These measurements complement
the long-baseline VLTI data by adding in visibilities at shorter baselines
up to 9 m. 
\section{Results}
\subsection{The WC stars}
The two WC type stars in our sample are WR 95 and WR 106. They are both
classified as late type (WC 9) and are known to be persistent dust producers
from IR excess detected over long-term photometry \citep{wil95}. 
The photometric data do not have any periodicity and neither do the spectra 
indicate any clear signs of binarity. 
We obtained two MIDI visibility measurements
on WR 95 on consecutive nights in July 2004. Though the projected baseline
and position angle differ slightly between these two measurements,
the visibility values were remarkably similar (Figure 1, broken lines). Though this might indicate a
small degree of asymmetry, the accuracy of these measurements are not high
enough to justify this interpretation. Instead we have chosen the average of 
the two sets to represent the spectrally dispersed visibility (Figure 1, solid line). 
As can be seen in Figure 1, we clearly resolve the WR 95 dust shell with 
visibility values ranging from $\sim$ 0.2 to $\sim$ 0.6 over the 
8-13 \micron\ band. In the absence of detailed knowledge of the source geometry
we have assumed a gaussian profile and the resulting FWHM in milliarcseconds (versus wavelength)
is shown by the dotted line. WR 95 was not observed with the Keck segment-tilting method 
since it was below the sensitivity limit.  

Figure 2 shows the MIDI results for WR 106. Though
we obtained two measurements in July 2004 (see Table 1), poor seeing conditions
resulted in variations in fringe tracking and calibration between the two
observations. In the absence of a clear criterion to select one data set
over the other, we average the two and the resulting larger error bars are
shown in the figure. However, this target
is well-resolved and meaningful conclusions can be drawn even from the
less-than-optimum data. The visibility variation with wavelength is similar
to that for WR 95. 

WR 106 was resolved 
by the Keck single aperture (Table 2). This was surprising in that, given the visibility 
values measured by the VLTI, one would have expected an unresolved source 
at the comparatively short baselines offered by the Keck mirror.  
This indicates that there is an extended 
component in the dust distribution which was over-resolved and hence failed
to show up in the VLTI
observations. We present these results in  Section 4 where
we discuss some simple models.

In Table 3, we list the angular sizes (gaussian FWHM) measured with MIDI at 10.5 \micron\ 
and the corresponding
linear sizes of the objects, assuming 
distances from the literature (see table footnote). 
The parameters directly constrained by this
data set are the size of the dust shells and its variation with
wavelength. The radius at which the dust grains start to condense is a much-debated
and crucial topic in the theory of dust-formation around WR stars 
\citep{zub98,cro03}. The major difficulty has been that
the equilibrium temperatures at the radii predicted by the SED-based models
are often above the sublimation temperature for carbon dust. 
Dust formation in the shock interface of a colliding-wind system
has found some favor in this context. A recent summary of results from
a long-term
campaign of aperture-masking interferometry at the Keck \citep{mon07}
provides the near-IR sizes for a number of WR stars, including those in
our sample. The authors point to a tight correlation between the
the surface-brightness in the 1.5 - 3.0 \micron\ band and the sizes (gaussian
FWHM) of the objects in their sample to argue for a common underlying 
dust-production mechanism.
Our direct measurement of {\em the mid-IR} size-scale for these
objects is an
important addition to this debate. 
From our gaussian FWHM curves, it is clear that the
size of the emitting region is more or less invariant in the 8 - 13 \micron\ band. 
This could indicate that the emission is dominated by an inner
edge or ring of material. 
In fact the sizes for WR 95 at the shorter wavelengths (1.6 and 2.2 \micron) reported 
in \citet{mon07} are quite similar to our values. For WR 106, the mid-IR size
is larger, as expected from the presence of the extension seen in the Keck measurement.
In section 4, we present radiative transfer
models for the dust shells based on our measurements and derive 
inner edge radii 
which we compare with values from SED-based models.

\subsection{NaSt1}
NaSt1 has a checkered history of classification. It was initially typed as
a late-type WN star, WR 122 \citep{vdh81}, but subsequently this classification
has been questioned. The
most recent study \citep{cro99} has resulted in this star not
being considered a Wolf-Rayet any longer. Crowther and Smith find the chemical
composition of the optical nebulosity sorrounding the star to be similar in
many respects to an LBV, like $\eta$ Carina, and conjecture that the obscured
central star could be an early type WN, presumably just leaving the LBV phase.
The uncertainty in classification is in part due to
the rarity of these very massive stars, especially one caught in transition.
As such our measurement of the extent of the close-in dust and its
variation with wavelength is of significance.
The source is unresolved by MIDI in single dish imaging mode and the flux measured within the $\sim$300 mas beam is similar to the the value quoted in Smith and Houck (2001, Figure  4 in their paper) from a bigger beam, indicating that all the  mid-IR emission 
originates from a compact object.  Another notable feature is the absence (as in the WCs) of any silicate
feature in the visibility profile (Figure 3) or the MIDI spectrum (Figure 4).
Although this does not rule
out the presence of silicate dust, it attests to the fact that there are no structures with silicate emission at the MIDI resolution scale. This is in agreement with the finding of \citet{cro99} that the nebula is highly enriched in CNO processed material.

NaSt1 was not resolved with the Keck single aperture. In our experience,
the segment-tilting method is capable of resolving sizes down to  
40-50 mas at the 3-sigma level. At 10.7 \micron, this is approximately a factor of 6 better than 
the formal diffraction
limit (1.22 $\lambda$/D) for the 10 m aperture. A more rigorous analysis
for the resolution limit for aperture-masking observations at Keck 
(the near-infrared analogue of segment-tilting) estimates it to be 4 times better than the
diffraction limit \citep{mon07}. We estimate an upper limit of
50 mas for NaSt1 from the Keck experiment.

NaSt1 was observed with MIDI at both the 
UT2-UT3 (projected baseline: 44.0 m, PA: 42\degr)
and the longer UT2-UT4 (projected baseline: 63.0 m, PA: 78\degr).
Figure 3 shows the visibility values and corresponding gaussian sizes for NaSt1.
The visibility varies with wavelength,
decreasing towards the red end
of the spectrum. This typically indicates cooler material further away
from the star and is in contrast to the behavior we see in the
WC stars. At least part of the dust around LBVs could be swept from the
ISM and/or formed in a cooler environment in the
episodic ejecta from the LBV phase and could lead to a less-defined inner edge and 
more spread in the dust temperature. 
The visibility on the PA 42\degr\ shorter baseline is very similar to that on the longer one, 
and if we intrepret this
directly as a size, may indicate an elongation in this direction. The optical image of NaSt1 \citep{cro99}
in the NII $\lambda$6583 line
shows an elliptical nebula with the major axis at a PA of 30\degr and an aspect ratio of 0.6, though the size
scale ($\sim$ 8 arcsec) is several orders of magnitude larger than the
milliarcsecond scales we measure.

However, it must be stressed that the visibility curves from the two different 
baselines are almost identical and the 
probability of obtaining the same profile for a spatially complex object (even if as 
simple as an ellipse) by pure chance is 
very low.
A possible alternative interpretation is that, following the 
suggestion of \citet{cro99}, the mid-IR 
source could be split into two components: a compact, unresolved part, with a correlated flux of 
about 5~Jy from 8 to 13 \micron\
that corresponds to the central source and perhaps some hot dust in its immediate surrounding, 
and a detached 
$\eta$ Car-like nebula, fully resolved by the interferometer. The extension of this nebula may be of the same order as 
that of the region contributing the [NeII] 12.8 \micron\ line which is seen in the single dish spectrum, but not in the 
correlated (i.e., arising in the unresolved core) flux (Figure 4). However, the upper limit 
on the size
of this component would be limited to $\sim$ 50-60 mas at most; anything bigger would 
have been resolved by the
Keck observations. Given the 10\% error on the MIDI visibilities, this could still be
over-resolved, though marginally, even at the shorter 44 m baseline. 
The size of the compact component is possibly less than 5 mas or so,
given that higher values would start to affect the relative slope of the visibilities
between the two baselines. 
Qualitatively, this hypothesis offers a natural explanation of the measurements: 
the central source is unresolved and the extension 
overresolved whatever the baseline length (44 m or 63 m). However, the
tight constraints on the extended component weaken the case. Clearly,
more observations are needed to clarify the picture.  
A very hot source (T$\sim$100,000 K) is required for the fitting the object's SED 
by \citet{cro99}, and the 5 Jy (1.5 $\times$ 10$^{-13}$ Wm$^{-2}\mu$m$^{-1}$) correlated flux we measure at 10 $\mu$m 
is not consistent with this. A constant 
correlated flux from 8 to 13$\mu$m (Figure  4) can hardly be attributed to a source at a single 
temperature (say, $\sim$ 2000 K) either. In order to 
decompose the SED into multiple sources at different temperatures it is 
important to study this object in the near-IR using the AMBER instrument 
at the VLTI to resolve the K band 
continuum and the very strong HeI 2.058 \micron\ line that may 
originate in the close vicinity of the 
star \citep[their Figure 6]{cro99}.

At a distance of 2 kpc, the maximum linear size from
the gaussian model is $\sim$ 40 AU and with a two-component model it is $\sim$ 100 AU (for comparison, 
the average extension of the optical nebula is $\sim$ 13600 AU at 2 kpc from \citeauthor{cro99}). If we take the same 
characteristic expansion velocity  as
\citeauthor{cro99} (15kms$^{-1}$) then the dynamical age of the dusty 
nebula at 100 AU from the source is about 30 yr. 
This shortens to a couple of years if instead we assume dust grains forming in the wind at a velocity of $\sim$ 100 -300 
kms$^{-1}$\citep{zub98,cro99}.  
Hence the dust that we resolve points to very recent or ongoing dust formation close to the star which has implications for its 
current poorly-known evolutionary state. 

\subsection{AG Car}
AG Car is one of the best studied LBVs and a prototype of its class. 
AG Car has been extensively observed from UV to radio wavelengths with photometric, spectroscopic, and polarimetric techniques 
over several decades \citep{hum89}, which locates it in the H-R diagram 
very close to the Humphreys-Davidson limit, an empirical luminosity boundary 
for evolved massive stars \citep{hum79}.

AG Car is surrounded by a bright extended bipolar nebula, 30" $\times$ 40" 
in size \citep{not92}, perhaps from a massive 
erruptive event 
10$^4$ years ago \citep{lam01}. The kinematics of the bipolar nebula is consistent with a two-lobe polar ejection viewed nearly from 
the equator \citep{not92}. Spectroscopic analysis by \citet{gro06} 
supports the bi-polar geometry and shows the 50-100 R$_\sun$ supergiant 
central star to be rotating
close to its break-up velocity. 
The high rotational velocity after the huge LBV outburst may be understood if the original explosion expelled the mass and momentum 
preferentially toward the poles. The wind at smaller latitude may not have formed or may have launched material with velocities lower than 
the escape velocity which fell back onto the star. 
The nebula contains a large amount of dust 
\citep[$\sim$0.2 M$_\sun$]{voo00}, which implies a 
huge mass loss rate during the eruption (at least 8-15 M$_\sun$).

Our measurements are set in context by two studies in particular; \citet{voo00} and \citet{not02}.
The images in \citet{voo00} obtained with the 10 \micron\ TIMMI camera on the 3.6 m NTT at La Silla 
reveal a central peak and a detached nebula with a radius 
of $\sim$ 10\arcsec. Their SED analysis
reveals a complex situation: 
the dust is 
dominantly oxygen-rich but Polycyclic Aromatic Hydrocarbon (PAH) features are 
observed at 3.3, 7.7, 8.6 and 11.3 \micron\ revealing some amount of 
carbon dust. 
Moreover, a population of very large grains ($>$ 10\micron) is 
necessary to explain the flux levels at long wavelengths whereas a 
population of small warm grains is invoked to explain the flux between 5 and 20\micron.  

There is evidence for CO J = 1$\to$0 and 2$\to$1 emission associated with AG Car \citep{not02} and the authors conclude that the
unresolved CO 
emission must arise close to the star in a high density warm region and not in the detached nebulae resolved in optical 
emission-line images. They postulate that a warm and dense equatorial disk exists close to the star.

MIDI observations of AG Car were carried out with the UT2-UT3 baseline on two nights, with projected baselines of 42.0 m and 35.6 m and
PA of 41\degr\ and 69\degr\ respectively. The visibilites and corresponding gaussian sizes are
shown in Figure 5 and the spectrum in Figure 6. AG Car is marginally resolved. The measured size ($\sim$ 11 mas) is similar
for both observations and 
fairly uniform with wavelength. 
The \citet{not02} optical images show strongly enhanced H$\alpha$ 
brightness at PA$\sim$35-225\degr, and the morphology is similar 
at 12$\mu$m \citep{voo00}. This direction is also the waist of the nebula, the largest extension being seen perpendicular to it.
The MIDI baselines are therefore oriented close to an optimal direction to detect equatorial material and we are 
very likely resolving the equatorial disk, though we cannot presently distinguish between a disk and shell structure. 
This is the first 
direct evidence for dust in a compact structure. 
The size we detect (11 mas) is $\sim$ 3 orders of magnitude smaller than the upper limit (3'') to the size of the (unresolved) CO outflow in 
\citet{not02}. 
The corresponding linear size is of
the order of only 100 AU at a distance of 6 kpc; very close to the central star. Nevertheless, this detection adds  
weight to the \citet{not02} claim that the CO emission arises from molecular gas close to the star and is shielded by a disk structure. 
The presence of dust has implications also for the origin of this compact structure. \Citet{not02} favor a scenario where the
the star goes through a Red Super Giant (RSG) or RSG-like phase \citep{smi98,lam01} during which the molecular outflow is formed. 
Dust formation is also commonly observed during the RSG stage \citep{mas05,dan94} and fits in with this hypothesis, 
though the $\sim$ 100 AU scale size indicates very recent or current dust formation as well.

There are few published spectra of AG Car in the mid-IR and the MIDI spectrum is a valuable addition. 
For comparison we also
show in Figure 6 the spectrum from the Short Wavelength Spectrometer (SWS) on the Infrared Space Observatory (ISO) 
in the 8-13 \micron\ range \citep{slo03}, scaled to roughly match the flux level of the MIDI spectrum.
The ISO spectrum, with a much bigger aperture, exhibits a slightly different slope from 
the MIDI measurement, but the striking difference is that the MIDI spectrum does not show the 11.3\micron\ PAH feature, implying that 
the PAH originates from further out in the disk/shell as conjectured by \citeauthor{voo00}.   

\section{The WC Models}
The dust around late type WR stars has been closely studied over the last
two decades or so and presents a challenging problem pertaining to
its formation, geometry and evolution. There have been quite a few
attempts to model the dust in some detail. WvdHT 
model these as spherical shells based on extensive IR
spectroscopy. \citet{zub98} provides dust
models with emphasis on the physics of grain formation and growth 
taking into account
the grain dynamics in the stellar wind. \citet{har04} take advantage of
the high resolution near-IR images of WR 104 \citep{tut99}, which show a ``pinwheel'' 
of dust formed in a binary wind interface, to construct a 3D radiative transfer model for this
object. As mentioned earlier, a large percentage of the known dust-forming
late type WC stars are known to be binaries, WR 104 being a prime example. 
The formation of dust in
these stars is usually interpreted using the wind-wind collision model
of \citet{uso91}. None of the stars in our sample show definite evidence  
of being  binaries \citep{wil00}, though this remains a strong possibility. 

It has been difficult to explain dust 
formation in the hostile environment of the wind from single WR stars.
SED-based models typically have dust forming at distances from 
300 up to a 1000 times the stellar radius \citep{zub98}. The 
consensus on the material of the dust, from the SED signature, is amorphous
carbon (WvdHT). However, at these
distances the electron temperatures in the winds of WRs are likely as high as
10$^4$ K \citep{zub98}, implying that carbon is almost fully 
ionised. 

We present here 
radiative transfer models for the WC stars in our sample with an aim
to use the interferometric measurements to answer the crucial question as
formation radius of the inner edges of these dust shells.  
To this end, we use a 2D Monte Carlo radiative transfer code, MCTRANSF \citep{nic03},
to construct spherically
symmetric dust distributions consistent with the measured visibilities.  

The parameters of the models are the star diameter and effective temperature, the
internal and external radius of the envelope, its density profile and 
optical depth, the minimal and maximal radius and size distribution of the 
grains and the optical indices of
the dust. 
We start with the basic parameters for the
dust shells around WR 95 and WR 106 mostly taken from the comprehensive 
SED-based models of WvdHT (see Table 4). 
By propagating energy packets in the sampled envelope (circular grid), 
taking into account the absorption, emission and scattering characteristics 
of the dust, MCTRANSF computes the temperature of the envelope, the SED of 
the flux received by the observer directly from the star, as well as the flux 
emitted and 
scattered by the envelope. 
The code has been described in detail in \citet{nic03} (see
\citeauthor{pas04}, \citeyear{pas04}, for benchmarking tests). 
An example application in
modeling a circumstellar dust shell is found in \citet{woi05}.

We then explore the parameter
space around this model by primarily varying the location of the inner edge of
the disk to see what values are most consistent with both our measured 
visibilities and spectra. We also tried limited variations of the optical depth, 
grain size and
distribution to quantify their effect on the visibility profile and spectra.
The radius of the central star and the composition of the dust were not
changed in these trials. 
MCTRANSF does not explicitly incorporate the sublimation temperature of the 
grain as an input.  The dust temperature in the nucleation zone 
from existing observational and modeling constraints ranges from 1000 - 1500 K 
or so \citep{zub98, har04} and in the process of iteration we discard those
models which produce untenably high temperatures at the inner edge of the shell.
We stress that the modeling is not a ``fitting'' exercise. That does not
seem justified, given 
the limited visibility data and the fact that the actual geometry of the dust is
most likely more complex than simple spherical shell models. Our goal here
is to use the new measurements
to constrain the vitally important distance to the dust-forming
zone from the star. 
 
\subsection{WR 106}
Extensive studies of the SED of this well known dust-producing WC9 star exist
in the literature and long-term IR photometry is summarised in \citet{vdh01b}.
Infrared (J-Q bands) measurements are presented in
WvdHT.  More recent mid-IR ISO spectra are found in \citet{smi01}.     
Given the difficulty in standardising mid-IR photometry from
various sources and the fact that dusty WC stars can have differing
degrees of variability, we rely mostly on the 8-13 \micron\ 
spectra obtained with the MIDI instrument simultaneously with the
visibilities. \citet{lei04} have confirmed the reliability
of spectra from MIDI. We used standard reduction techniques as
they describe to extract the spectra; most of our interferometric 
calibrators were also chosen to be IR spectroscopic standards
\footnote{\url[HREF]{http://ssc.spitzer.caltech.edu/irs/calib/templ/cohen\_models/}} and 
as such
provide ideal templates to calibrate the spectra. 
De-reddening WR spectra has been non-trivial, primarily because these stars are
grouped towards the Galactic center with large columns of intervening
interstellar dust in addition to the circumstellar dust. The wide interstellar silicate feature centered 
around 9.7 \micron\ complicates the issue \citep{smi01}. To 
provide a uniform comparison, we have de-reddened the 8-13 \micron\
band with values of A$_\lambda$/A$_{\rm v}$ from WvdHT and the van der Hulst
extinction curve 15 for the shorter wavelengths. 

For WR 106, the model visibilities from our ``best'' model are shown in Figure 7, 
along with the Keck measurements as well as the MIDI visibility at 10.5 \micron.

The MIDI visibilities across the entire band (8-13 \micron) are compared
with the model in Figure 8. 
The closure phases measured by the Keck segment-tilting experiment did not
show any significant deviation from zero, signifying a lack of asymmetry at 
this resolution and sensitivity. We therefore plot the Keck visibilities 
treating the baselines as scalars (i.e., ignoring the orientation). 
Figure 9 shows the model SED along with our measurements and those from 
the literature\footnote{From catalogs available at the {\it VizieR} online database \citep{och00}. Optical fluxes: Homogeneous Means in 
the UBV System (Mermilliod, 1991); Hipparcos Input Catalogue, Version 2 (Turon et al., 1993); The USNO-B1.0 Catalog 
(Monet et al., 2003); IR fluxes: Catalog of Infrared Observations, Edition 5 (Gezari et al., 1999); Third release of DENIS (DENIS consortium, 2005).} 
spanning wavelengths from the optical to the far-infrared for comparison. 
The temperature profile of the dust in our model is shown in Figure 10. 
The parameters for the model are as below:

\subsubsection*{Central Star}
For the central star, we have adopted a blackbody spectrum at the
effective temperature listed below. The temperature of the
WR star has been assigned a range of values in the literature.
Most dust-formation models adopt a value between 18000 and 25000 K
(e.g., WvdHT). These values
have been criticized as being under-estimates from recent 
studies of the WR spectra \citep{cro03}. However,
we find in our models that higher temperatures (we tested up to 40000 K)
do not fit the 8-13 \micron\ spectra and also lead to  dust temperatures 
above the 1500 K or so sublimation temperature usually adopted for amorphous 
carbon grains.
This indeed has been
the reason that previous modeling attempts favour the lower values as 
well.
Although more sophisticated 
SEDs have been used for the stellar photosphere \citep[WdvHT]{har04}, 
a simple blackbody
is likely adequate for our purpose. The radius of the star is
is assumed to be 14.6R$_\sun$ (WdvHT) at a distance of 
2.09 Kpc \citep{vdh01a}.
\begin{itemize}
\item[*]Temperature of central star (T$_{eff}$) = 19000 K
\item[*]Angular radius of central star (R$_*$) = 3.23$\times$10$^{-5}$ arcsecond
\end{itemize}
\subsubsection*{Dust Shell physical properties}
\begin{itemize}
\item[*]Composition: Amorphous carbon (optical constants from 
\citeauthor{rol91}, 1991)
\item[*]Grain size: Ranges from 0.5 to 0.6 \micron
\item[*]Optical depth ($\tau$) = 0.015 at 10 \micron
\end{itemize}
\subsubsection*{Dust Shell geometry}
\begin{itemize}
\item[*]Inner radius of dust shell (R$_{int}$) = 280 R$_*$(multiples of stellar radius)
\item[*]Outer radius of dust shell = 8400 R$_*$
\item[*]Density profile: Proportional to  r$^{-2}$, 
except in the region between
1.0 R$_{int}$ and 14.0 R$_{int}$ where we introduce an over-dense
region with density proportional to 7$\times$r$^{-2}$. 
\end{itemize}
\subsection{Discussion}
Most of the starting parameters of the dust shell model were based on 
WvdHT.
They establish the composition to be amorphous C
at least in the case of the episodic dust-producer WR 140. The lack of 
silicate features in the mid-IR region also points to amorphous C.
The grain size and distribution for the shells are not well known. 
The \citet{zub98} grain-formation model predicts a fairly small maximum grain size($\sim$ 100 - 200{\AA}). 
\Citet{har04} model
the {\it inner part} of the WR 104 pinwheel nebula with small ($\sim$ 100{\AA}) grains, mainly to match the
2-10 \micron\ flux.
However, there is evidence in the SED of WR nebulae pointing towards large grains of $\sim$ 0.1 to 1
\micron\ \citep{chi01,vee98} and \citet{mar02} find a characteristic size of 0.5 \micron\ for the
extended emission in WR 112. 
The model presented here also favors large grains with a quasi-uniform 
grain size ranging from
0.5 to 0.6 \micron, with a standard interstellar distribution scaling 
as a$^{-3.5}$\citep{mat77}. 
A model with sizes ranging from 0.03 to 2.0 \micron 
also shows reasonable agreement with the data for similar values
of the other parameters. Despite
extensive trials we were unable to find a distribution of smaller grain sizes
which agreed with the data, in particular the flux at the shorter infrared wavelengths. 

The optical depth for these dust shells in the mid IR is fairly low 
\citep{zub98}.
We use here a value of 0.015. 
Our primary goal of matching the measured visibilities is not strongly
affected by the optical depth and it is the spectra that constrain this
parameter. 

The important constraint imposed by the visibility profile 
is on the range of {\em inner edge radii} of the dust shell. The ``best''
model in Figure 7 has a value of 280 R$_*$ for the inner edge radius, which 
is the same as in the WvdHT
model. We show in Figure 8 the model visibility profile
for  values ranging from 230 R$_*$ to 370 R$_*$.
Even for this narrow range, the 
visibility profiles are clearly different, 
indicating the fairly strong constraints imposed by the measurements.
    
The over-density (denser than the r$^{-2}$ profile) that we introduce in 
the shell is to accomodate the 
visibilities measured with the smaller baselines at the Keck,
which are lower than expected for a r$^{-2}$ profile consistent with the
longer baseline MIDI visibilities. The over-dense
region in the dust shell extends from the inner edge (R$_{int}$) up to 14 R$_{int}$ in 
the model,
i.e., size scales from 9 to 125 mas. This enhancement at the spatial
scales probed by the segment-tilting experiment (resolution $\sim$ 50 mas, see Section 3.2) 
decreases the visibility at the low spatial frequencies.
This of course is not a unique solution, but offers an
intuitive physical model.
The over-density could indicate an 
episode of more vigorous dust production. It could also be an artifact of
applying an overly simplistic spherically symmetric model to a possibly 
complex shell geometry as in, for example, a ``pin-wheel'' structure. 
In the absence
of more complete visibility sampling, we limit ourselves to simply
pointing out the existence of an extended structure in the dust around WR 106.      

\subsection {WR 95}
WR 95 is also a late-type dust-forming WC, similar to 
WR 106. As in the case of WR 106, there is no definite evidence so far,
either spectroscopic or photometric, of binarity. WR 95 falls below
the sensitivity limit of the Keck segment-tilting experiment and so
we base our models solely on the MIDI visibilities and spectra. Our best
model visibilities along with the measured values in the 8-13 \micron\ MIDI
band are shown in Figure 11.
Figure 12 shows the measured 8-13 \micron\  SED, the near-IR and far-IR photometry from the literature, and the model SED profile.
The dust temperature profile 
is as in Figure 13. 
The parameters of the model are as below:

\subsubsection*{Central Star}
\begin {itemize}  
\item[*] Effective temperature (T$_{eff}$): 25000 K
\item[*] Angular radius of star (R$_*$): 3.2$\times$10$^{-5}$ arcsecond
\end{itemize}
\subsubsection*{Dust shell physical properties}
\begin{itemize}
\item[*]Composition: Amorphous carbon
\item[*]Grain size: Ranges from 0.5 to 0.6 \micron
\item[*]Optical depth ($\tau$) = 0.005 at 10 \micron
\end{itemize}
\subsubsection*{Dust Shell geometry}
\begin{itemize}
\item[*]Inner radius of dust shell (R$_{int}$) = 480 R$_*$(multiples of stellar radius)
\item[*]Outer radius of dust shell = 1230 R$_*$
\item[*]Density profile: Proportional to  r$^{-2}$ 
\end{itemize}
As in the case of WR 106, we base our initial model parameters mostly on 
WvdHT. 
To match the near IR photometry and 
our visibility measurements we had to adopt higher
temperature of 25000 K for the WR star. The optical depth
for our best model, 0.005 at 10 \micron, is lower than that for WR 106. 
Since we do not have shorter baseline Keck
observations in this case to detect extended structure, a uniform r$^{-2}$
density profile suffices. The nominal inner radius of the shell is 480 R$_*$ (Figure 11 shows how the model visibilities
vary for slightly different values of this parameter), compared
to 410 R$_*$ derived by WvdHT. As in the case of WR 106, a dust grain distribution with sizes ranging from 0.3 to 2.0 \micron\ also yields a fair fit 
to the measurements.
\section{Conclusions}
High-resolution measurements of dust around massive stars are sparse in the
literature. Coupled with the rarity of the objects themselves, it is not
surprising that so much remains unclear about the mechanisms of dust-formation
and the nature of the dust itself. Only a handful of WR stars known
to produce dust have been imaged with any degree of detail. Three of these,
WR 104, WR 98a and WR 140 are binaries, producing dust at the wind-wind 
interface. The fourth, WR 112
reveals a multiple-shell dust structure, though at much 
larger (arcsecond) scales than discussed here for our targets, 
suggestive also of being a binary \citep{mar02}. Hence the measurements presented here, albeit
with limited baselines, of
similar objects with no previous evidence of binarity, 
are an important step in probing the nature of these
dust shells. 

We observed two WC9 stars (WR 106 and WR 95), an LBV (AG Car) and an object which may be 
in a stage of transition from an LBV to a WR phase, NaSt1. 
NaSt1 was well resolved and the
visibility measurements suggest increasing size with wavelength, a behavior 
unlike the WC 9 stars in our sample. We favor a
two-component model for the dust with a compact, unresolved
core and an extended component. Previous studies of the SED of this object have
also indicated this structure. For AG Car as well, our measurements resolve a compact
mid-IR source which is likely associated with a known molecular
gas component and is the first direct evidence for dust near the
central star. For both these objects, this study has revealed the presence
of close-in dust that has important implications for the state of evolution
of the central star.    

For the WC stars we construct spherically symmetric radiative transfer models
for the dust shells.
These models indicate the inner edge of these shells are at a few hundred
stellar radii (tens of AU) from the star.
Gaussian sizes (FWHM) directly estimated from the visibility 
measurement provide  model-independent upper limits which are consistent
with these values. 
At these distances, the dust temperature is fairly close to sublimation
values for carbonaceous dust. In addition, our results support previous
indications of fairly large grains ($\sim$ 1 \micron) despite these conditions.
These new results re-open an existing controversy: How does dust form in the 
hostile environment of a WR stellar wind at such high temperatures?  
The proximity of the inner edge to the star also forces us to adopt values
close to 20,000 K for the central star to keep the dust below the 
sublimation point. However, the spectra of late-type WC stars point to
higher (40,000 to 70,000 K) temperatures, rendering the situation even worse. 
Our
results certainly indicate that dust formation is likely occuring
under fairly extreme conditions. This may suggest a colliding-wind binary
model but direct evidence is still lacking for the objects in our sample. 

The dust around WR 106 seems to deviate from an r$^{-2}$ profile at
the shorter baseline Keck measurements. Our simple models suggest an over-dense
region fairly close to the inner edge. This could be also be a first indication of
of binarity with multiple shells from  episodic dust production or a more
complex geometry as in a pinwheel. But given
our limited baseline coverage, a caveat on modeling artifacts is warranted.

We have not attempted to estimate the dust mass from the models. Our 
measurements
are insensitive to the colder dust further from the star and it is unlikely
that we can improve on the existing dust mass estimates for these stars
from spectral profiles.

In the long-term, further progress would involve multiple baselines to image the dust 
structure. Observations of the outer dust at longer wavelengths are also required
to accurately estimate the dust mass and geometry. In the short-term, we aim to
look for asymmetries in the dust distribution by using the UT3-UT4
baseline at the VLTI, orthogonal to the one used here.





\acknowledgments

This work was performed in part under contract with the Jet Propulsion Laboratory (JPL)
funded by NASA through the Michelson Fellowship Program. JPL is managed for the
National Aeronautics and Space Administration (NASA) 
by the California Institute of Technology.
Some of the data presented herein were obtained at the W.M. Keck Observatory, 
which is operated as a scientific partnership among the California Institute of Technology, 
the University of California and NASA. 
The Observatory was made possible by the generous financial support of the W.M. Keck Foundation.
The authors wish to recognize and acknowledge the very significant cultural role and reverence 
that the summit of Mauna Kea has always had within the indigenous Hawaiian community. 
We are most fortunate to have the opportunity to conduct observations from this mountain.
We acknowledge the VLTI consortium for both technical and software support.  

We thank the referee for
comments which have significantly improved this article.



{\it Facilities:} \facility{Keck}, \facility{VLTI (MIDI)}

\clearpage
\thispagestyle{empty}
{\rotate
\begin{table}
\tabletypesize{\scriptsize}
\begin{center}
\caption{MIDI Observations \label{tbl-1}}
\begin{tabular}{cccccc}
\tableline
Star & V  & IRAS 12\micron & Type & Date Observed & Calibrators\\  
     &  mag  &    Jy            &      &            &             \\ 
\tableline
WR 95 & 13.4 & 4.56 & WC 9 & 07-30-2004 & HD 177716, HD 129456, HD 169916 \\
      &      &      &      & 07-31-2004 & HD 169916, HD 165135, HD 192947 \\
WR 106 & 11.93 & 11.67 & WC 9 & 07-09-2004 & HD 165135 \\
NaSt1 & 15.4 & 14.5 & WN(?) & 07-30-2004 & Same as WR 95 above \\
      &      &      &     & 09-29-2004\tablenotemark{a} & HD 168723 \\
      &      &      &     & 09-30-2004\tablenotemark{a} & HD 168723 \\
AG Car & 7.09 & 12.42 & LBV & 04-10-2004 & HD 81797, HD 107446, HD 129456, HD 139997\\
\tableline
\multicolumn{6}{l}{$^a$Baseline UT2-UT4}
\end{tabular}
\end{center}
\end{table}}



\clearpage

\begin{table}
\begin{center}
\caption{Keck Observations}
\begin{tabular}{ccc}
\tableline\tableline
Star & Date of Observation & gaussian FWHM\\ 
     &       &    mas (@ 10.7 \micron)  \\ 
\tableline \tableline
WR 106 &06-26-2005 & 56.0$\pm$6.9\tablenotemark{a} \\
NaSt1 & 06-26-2005 & 50.0 \tablenotemark{b} \\
\tableline
\end{tabular}
\tablenotetext{a}{Dominant error arises from calibration}
\tablenotetext{b}{Object was unresolved. Value is an estimated upper limit}
\end{center}
\end{table}

\clearpage

\begin{table}
\begin{center}
\caption{Measured Sizes (MIDI)}
\begin{tabular}{ccccc}
\tableline\tableline
Star & Distance & gaussian FWHM & Linear Size & References\\ 
     &   Kpc    &    mas (@ 10.5 \micron)    &      AU    &  \\ 
\tableline \tableline
WR 95 & 2.09& 28.4$\pm$1.5 & 59.4 & 1  \\
WR 106 & 2.3& 45.0$\pm$5.0 & 103.5  & 1 \\
NaSt1 & 2.0& 20.1, 14.8 ($\pm$2.0)\tablenotemark{a} & 40.2, 29.6  & 2 \\
AG Car & 6.0& 12.5$\pm$3.0 & 75 & 3 \\
\tableline
\end{tabular}
\tablenotetext{a}{The two values are from two different baselines. Interpreting sizes from visibility may be inaccurate for this object; see text}
\tablenotetext{}{References.-(1)\citet{vdh01a}; (2)\citet{cro99}; (3)\citet{hum89}}
\end{center}
\end{table}

\clearpage
\begin{table}
\begin{center}
\caption{Starting Parameters for the WC Models}
\begin{tabular}{lc}
\tableline\tableline
Parameter & Value\tablenotemark{a}\\
\tableline\tableline
T$_{{\rm eff}}$ & 1900 K \\
R$_*$ & 14.6 R$_{\sun}$ \\
Dust shell R$_{{\rm int}}$ & 410 R$_*$ (WR 95)\\
                           &280 R$_*$ (WR 106)\\
Dust shell thickness & 3 R$_{{\rm int}}$ (WR 95)\\
                     & 30 R$_{{\rm int}}$ (WR 106)\\
Dust composition & Amorphous C \\
Density Profile & r$^{-2}$ \\
Optical depth @ 10 \micron \tablenotemark{b}& 0.01 \\
Grain sizes\tablenotemark{b} & 0.1 to 1 \micron\\
\tableline
\end{tabular}
\tablenotetext{}{Unless stated otherwise, all values are from WvdHT}
\tablenotetext{a}{Single values are common for both stars}
\tablenotetext{b}{See text, Section 4.2, for references}
\end{center}
\end{table}




\clearpage


\begin{figure}
\plottwo{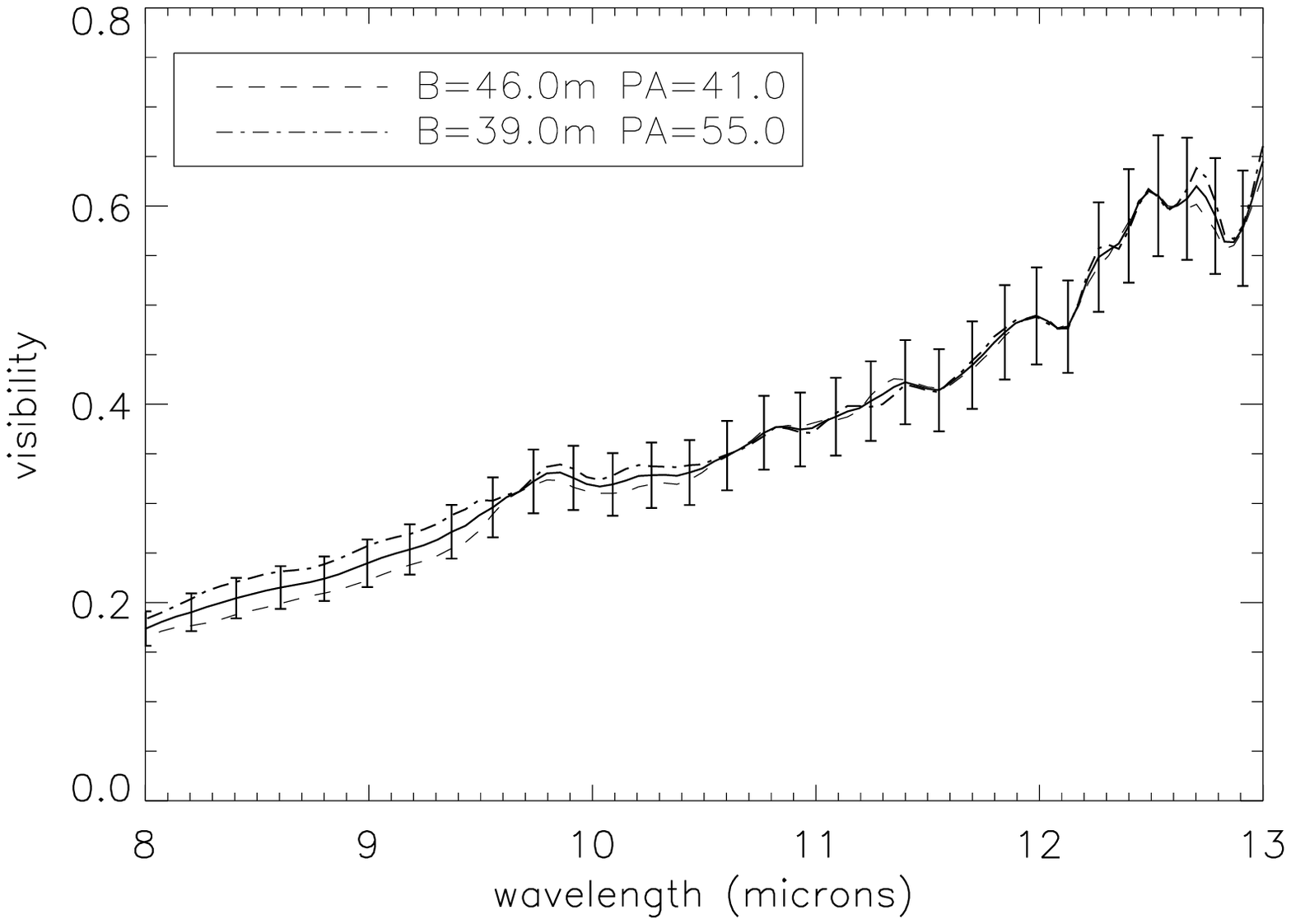}{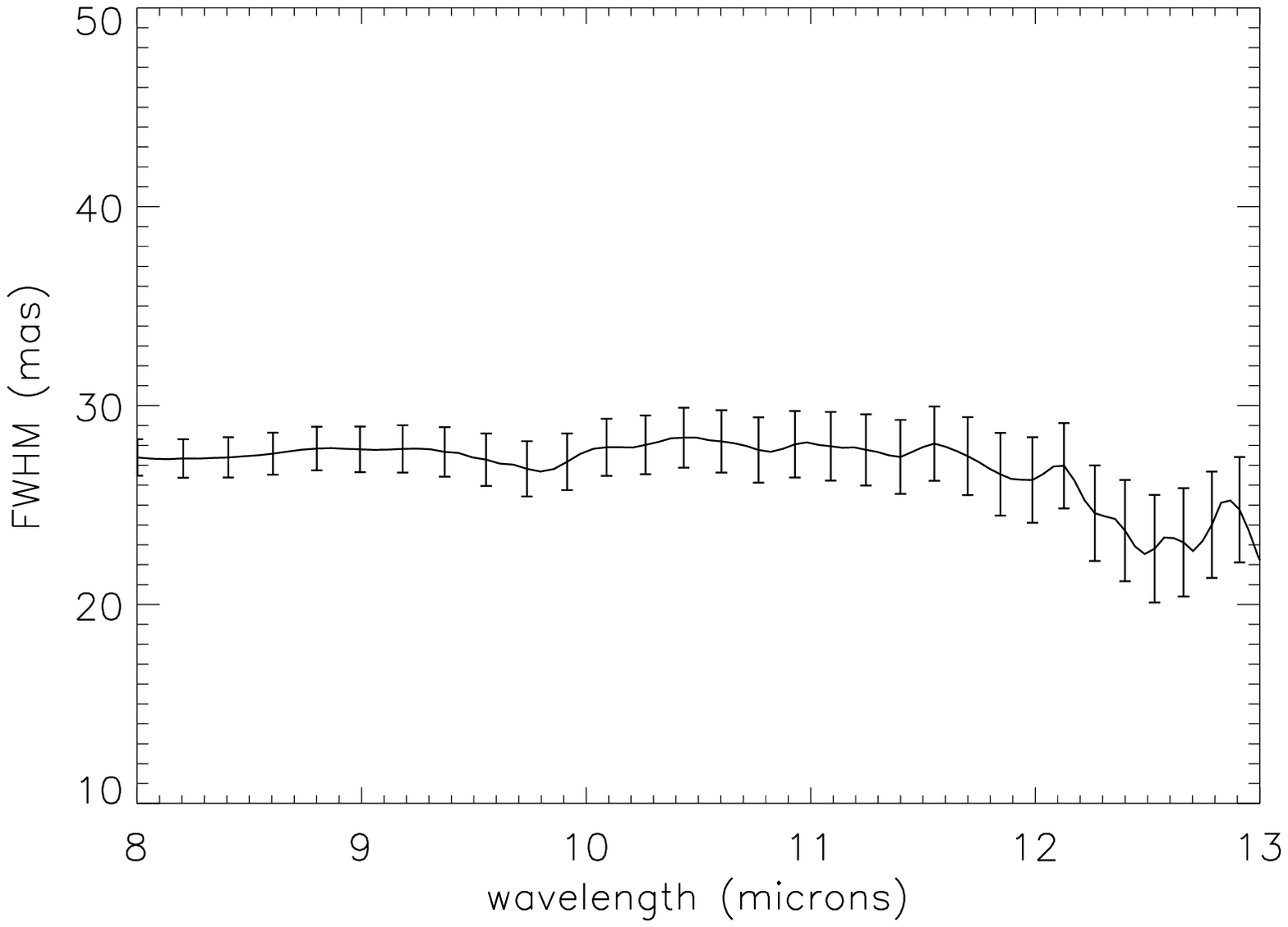}
\caption{Left: Visibility profile with wavelength for WR 95. The solid
line is the mean of the two measurements. 
Right: The corresponding (mean visibility) gaussian FWHM. \label{fig1}}
\end{figure}

\clearpage

\begin{figure}
\plottwo{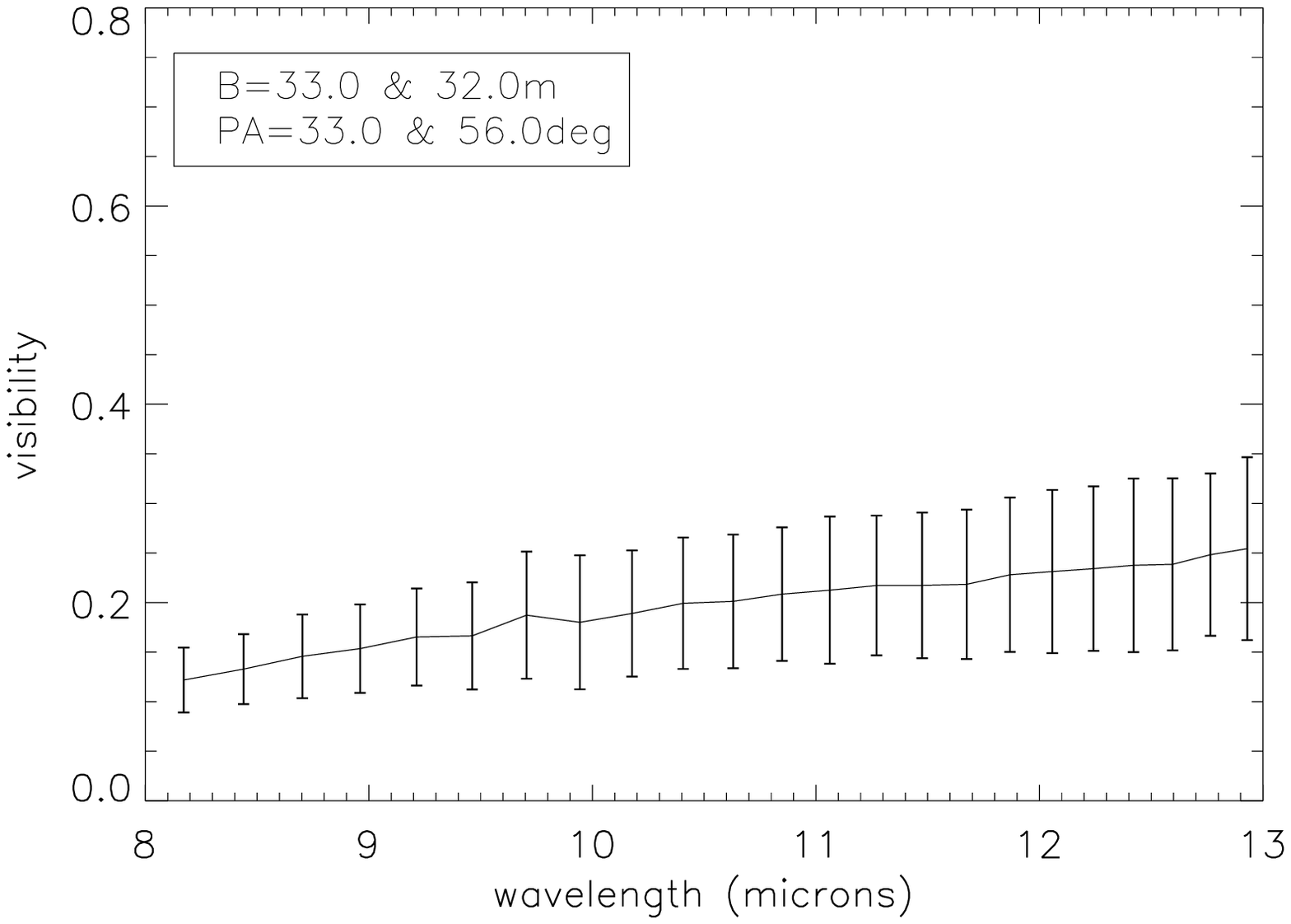}{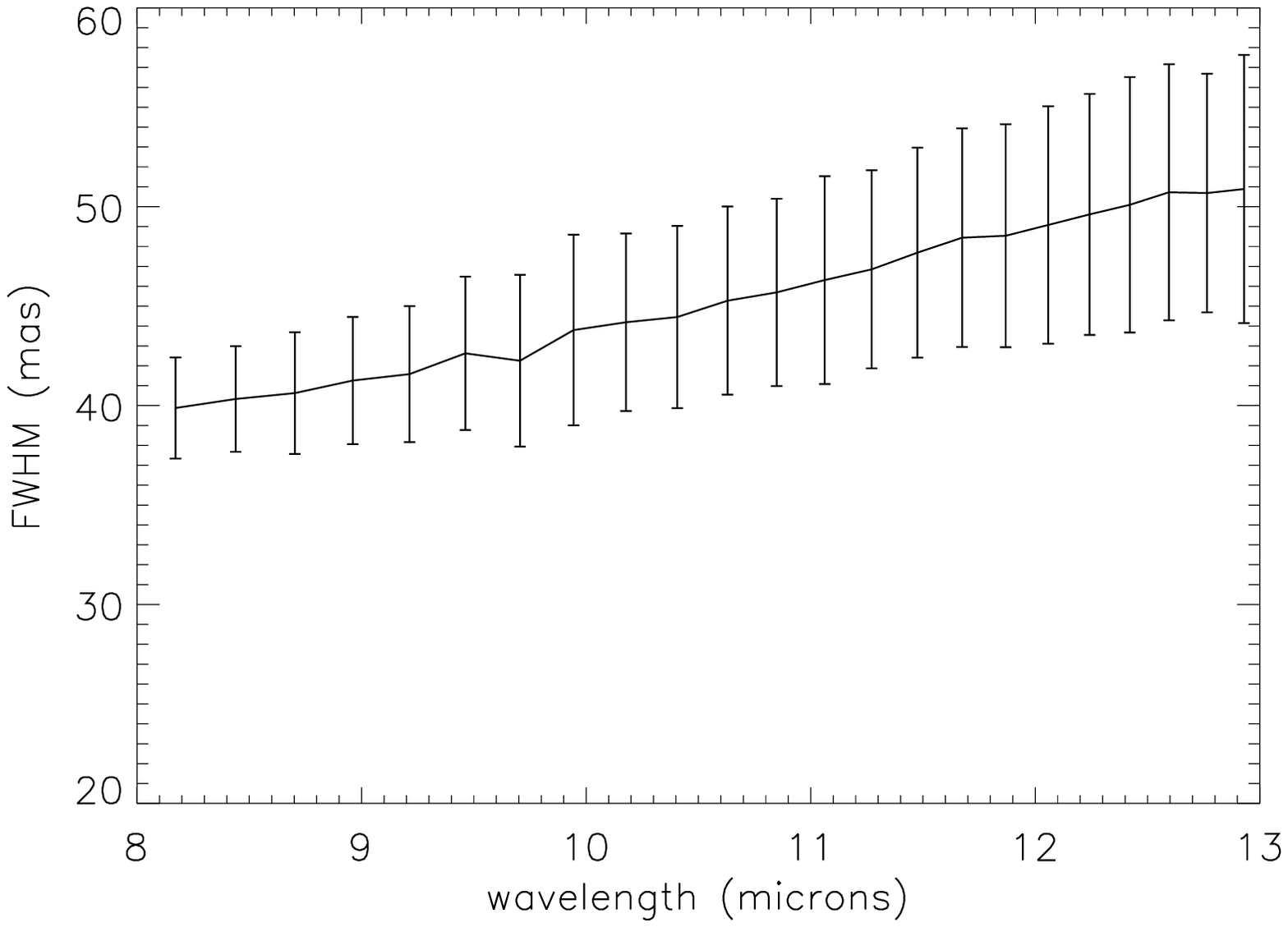}
\caption{Left: Visibility profile with wavelength for WR 106 
Right: The corresponding gaussian FWHM. \label{fig2}}
\end{figure}

\clearpage

\begin{figure}
\plottwo{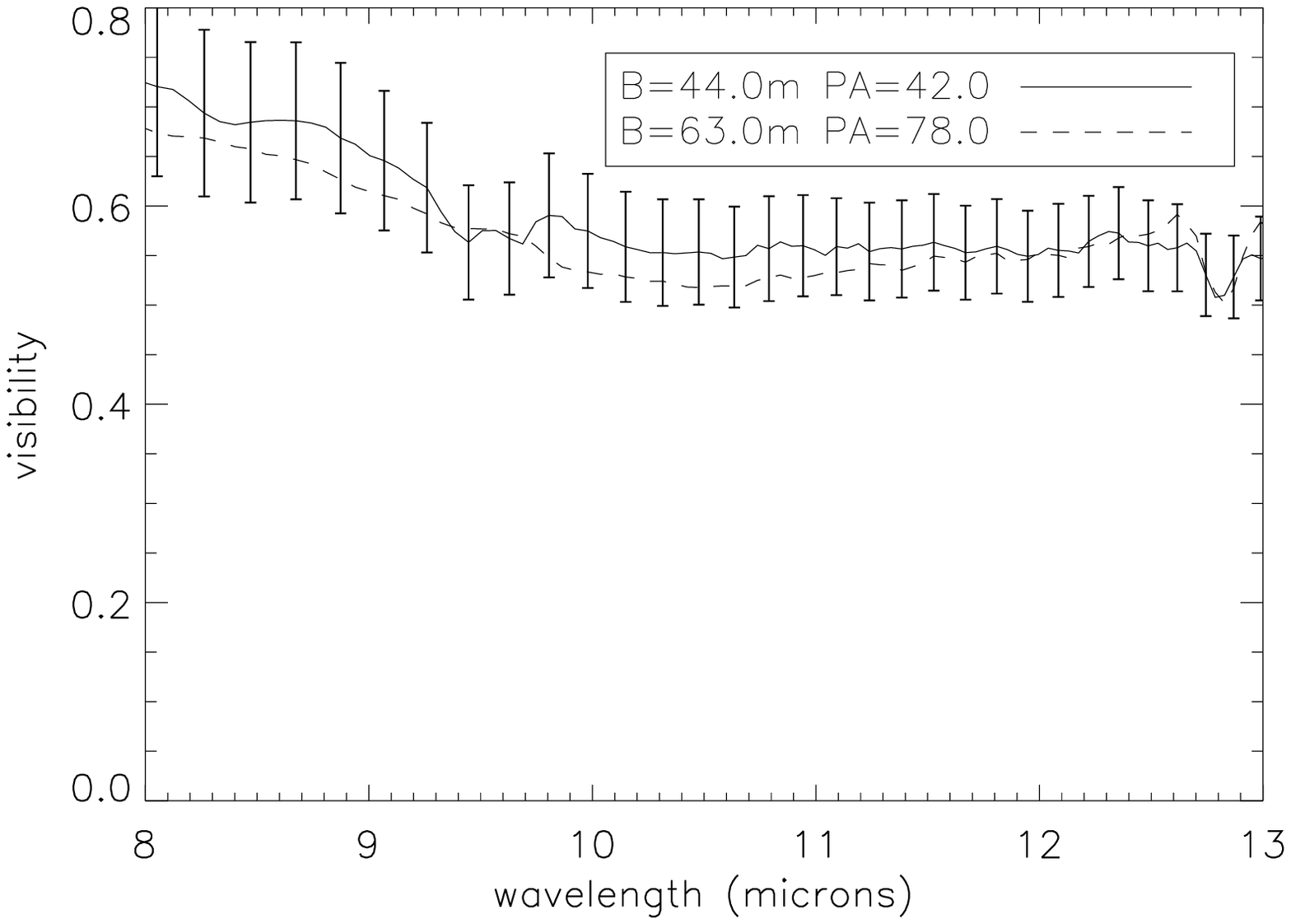}{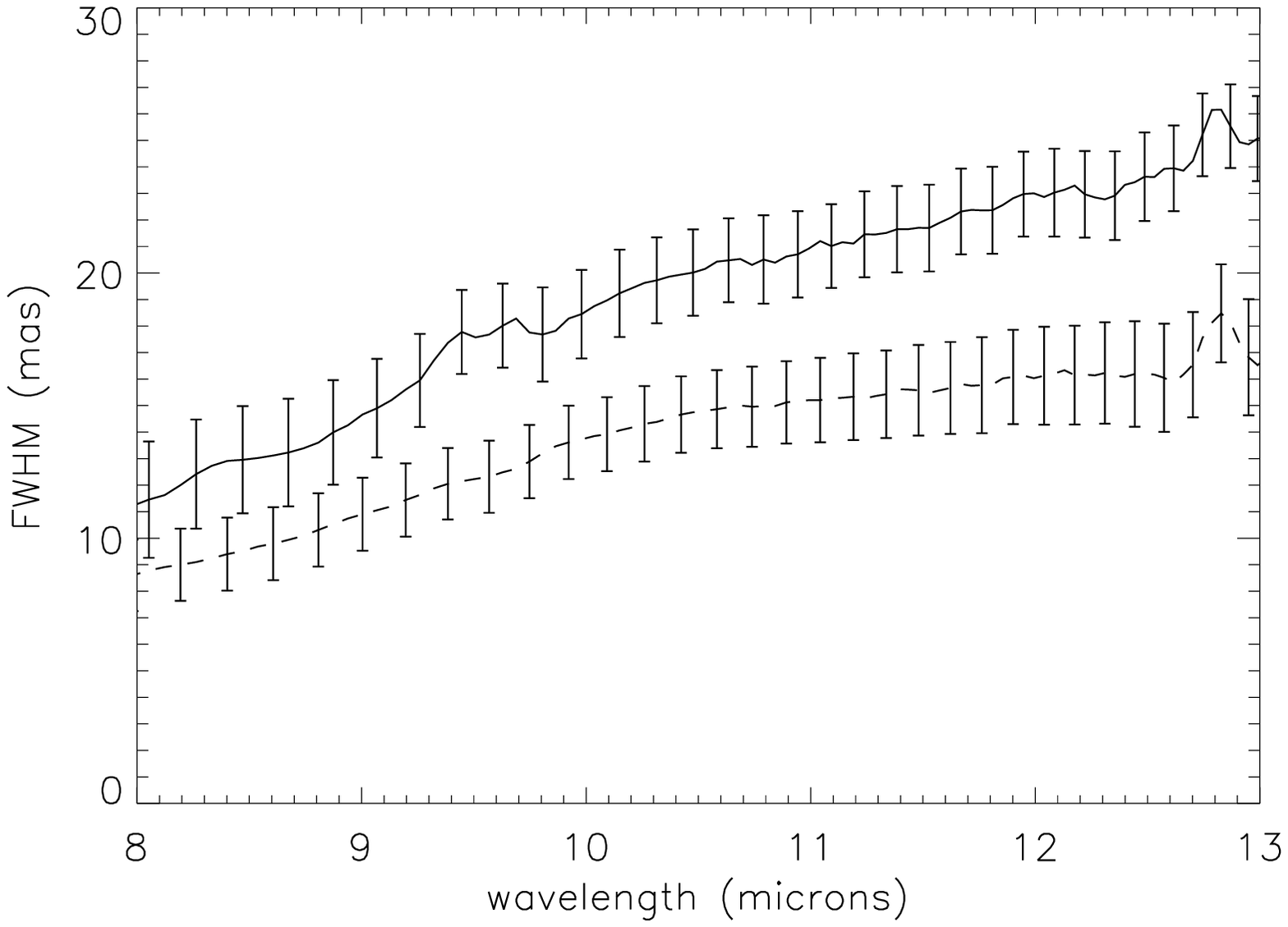}
\caption{Left: Visibility profile with wavelength for NaSt1 (only one set of error bars shown for clarity). 
Right: The corresponding gaussian FWHM. Transforming the visibility directly to a 
gaussian size is, however, probably inaccurate for this object (see text). \label{fig3}}
\end{figure}

\clearpage

\begin{figure}
\plotone{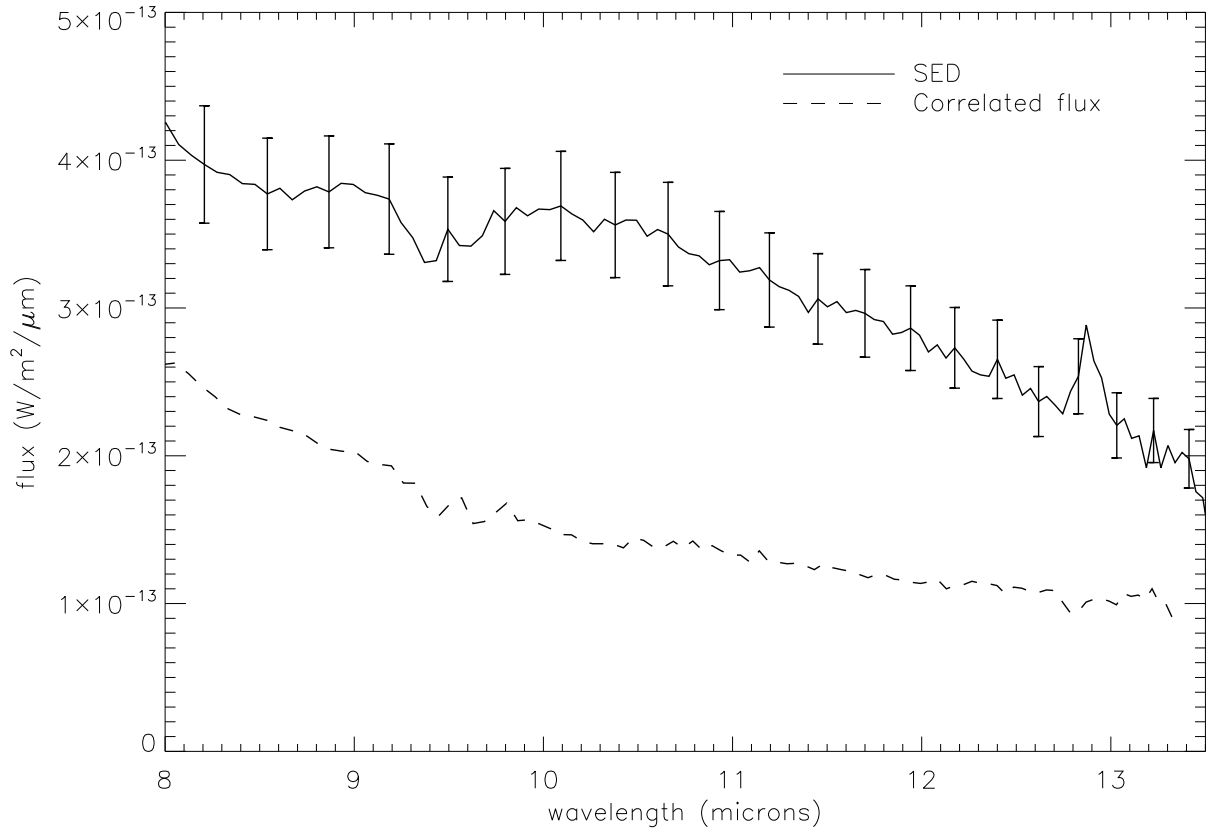}
\caption{MIDI spectrum and correlated flux for NaSt1. The Ne[II] 12.8 \micron\ feature is not seen in the correlated flux
and is likely fully resolved by the interferometer as is a large fraction ($\sim 1/2$) of the total continuum flux. 
\label{fig4}}
\end{figure}

\clearpage

\begin{figure}
\plottwo{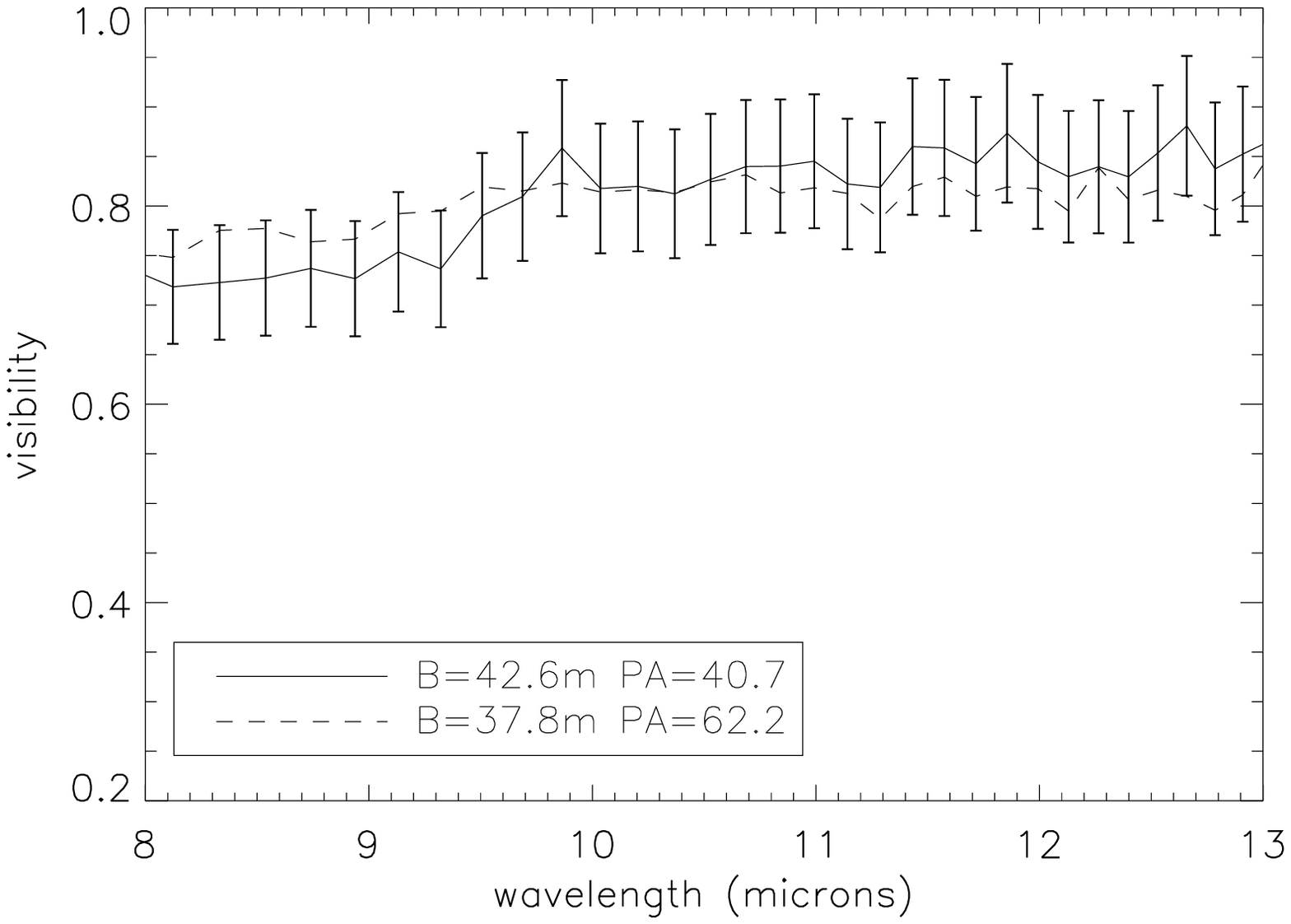}{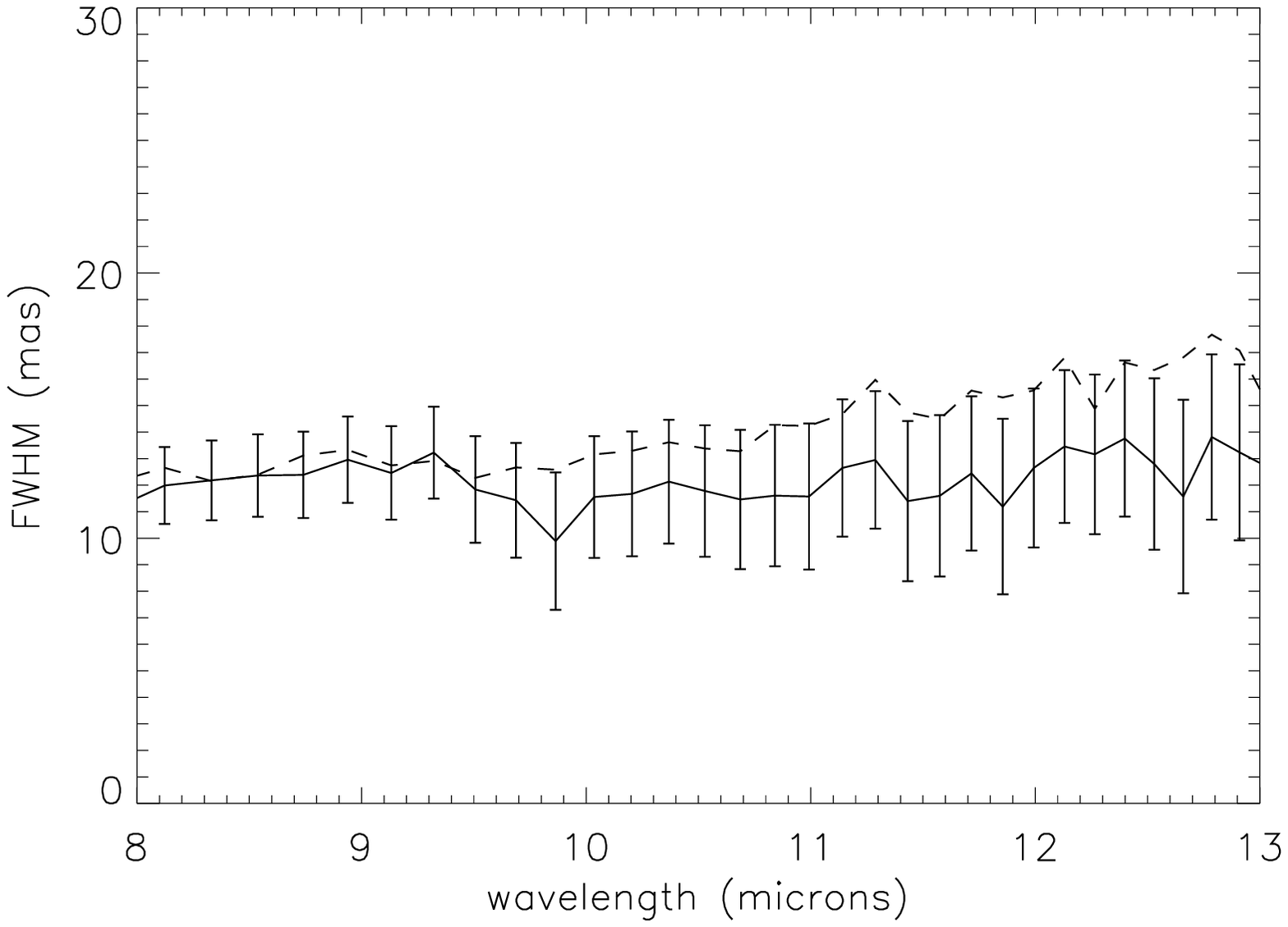}
\caption{Left: Visibility profile with wavelength for AG Car. 
Right: Corresponding gaussian FWHM. The error bars are similar for both measurements. For clarity only one set is shown.  \label{fig5}}
\end{figure}

\clearpage

\begin{figure}
\plotone{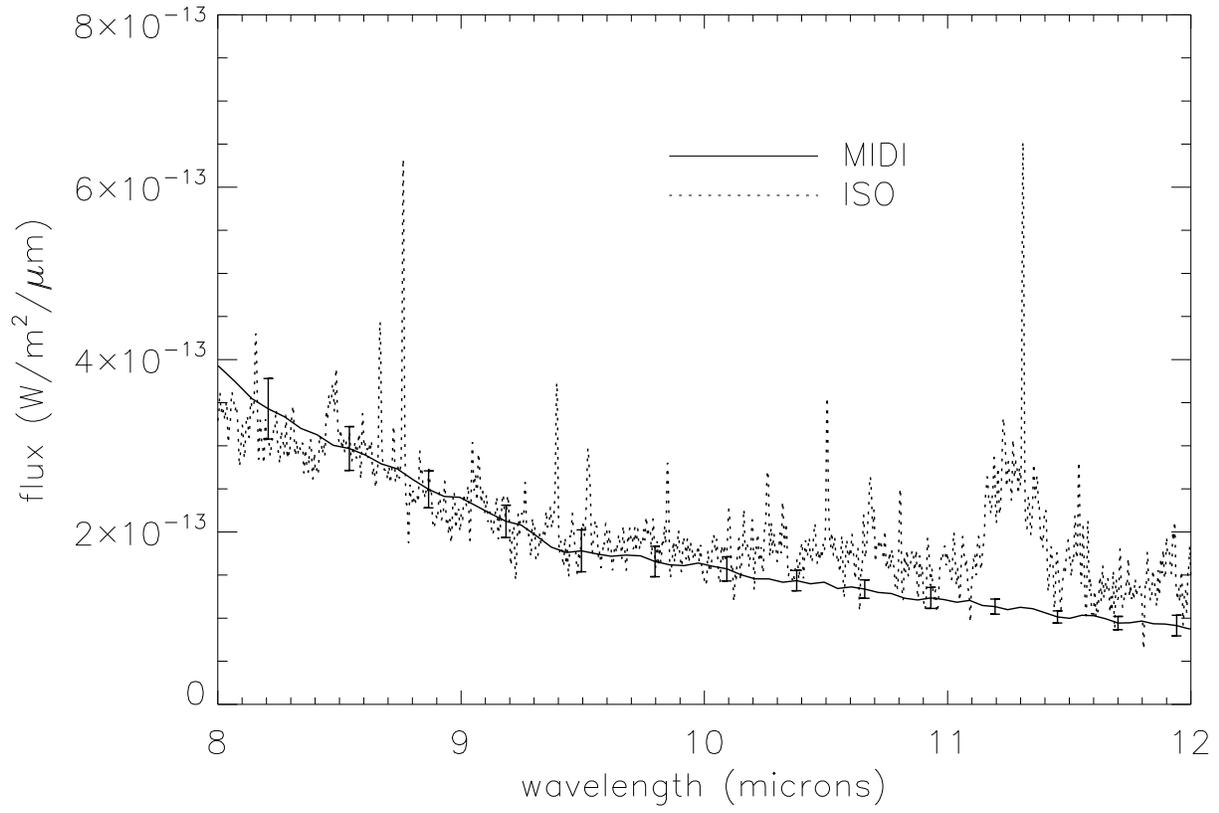}
\caption{MIDI spectrum for AG Car, with the ISO spectrum (scaled overall to match the MIDI flux level) for comparison. 
The PAH feature at 11.3 \micron\ is not seen in the MIDI spectrum.
 \label{fig6}}
\end{figure}

\clearpage

\begin{figure}
\plotone{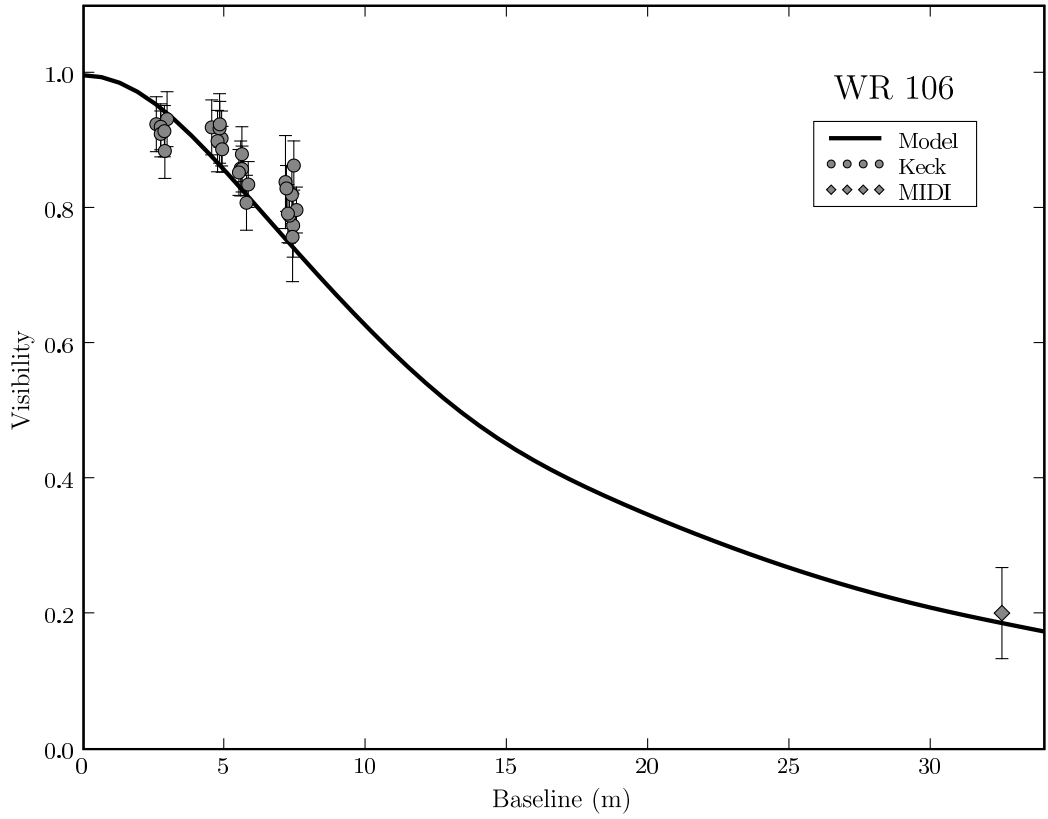}
\caption{Visibility versus baseline for WR 106. The solid line is the model visibility, the filled circles are the Keck visibilities (upper left) 
and the diamond, MIDI at 10.5\micron.\label{fig7}}
\end{figure}

\clearpage

\begin{figure}
\plotone{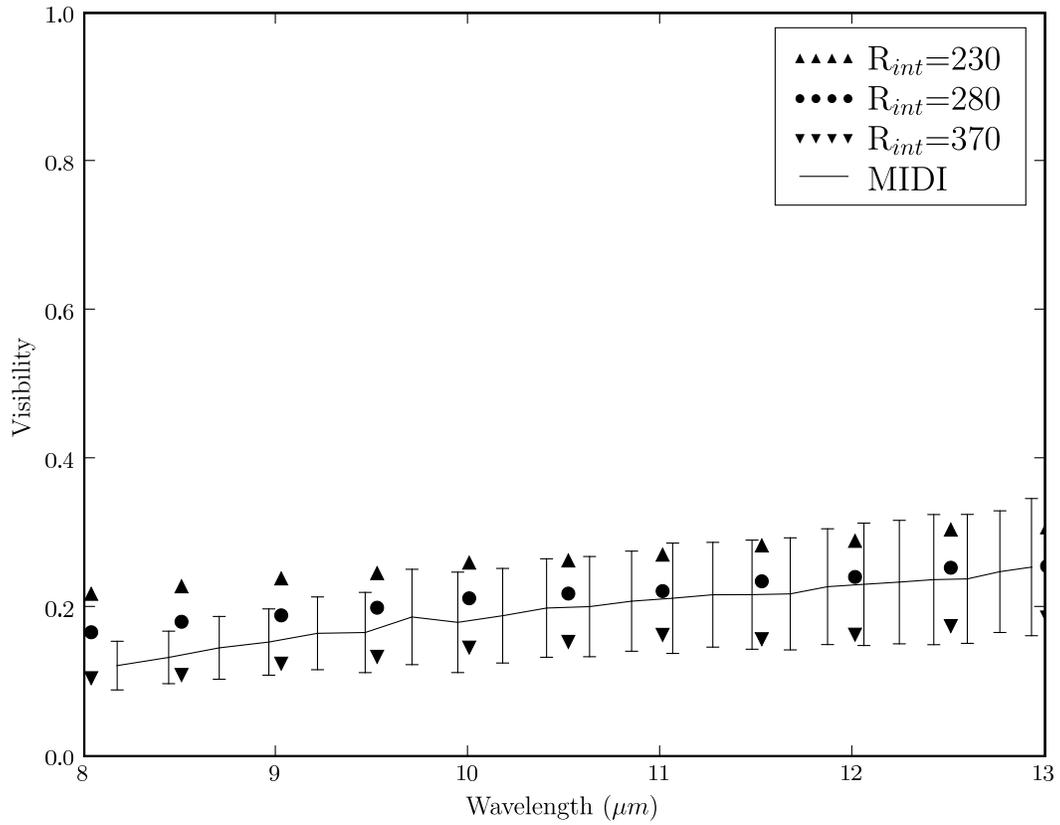}
\caption{Visibility versus wavelength for WR 106. The line (with error bars) is the MIDI measurement.
Models with varying inner radii, R$_{int}$ (in units of R$_*$), for the shell are shown for comparison.\label{fig8}}
\end{figure}

\clearpage

\begin{figure}
\plotone{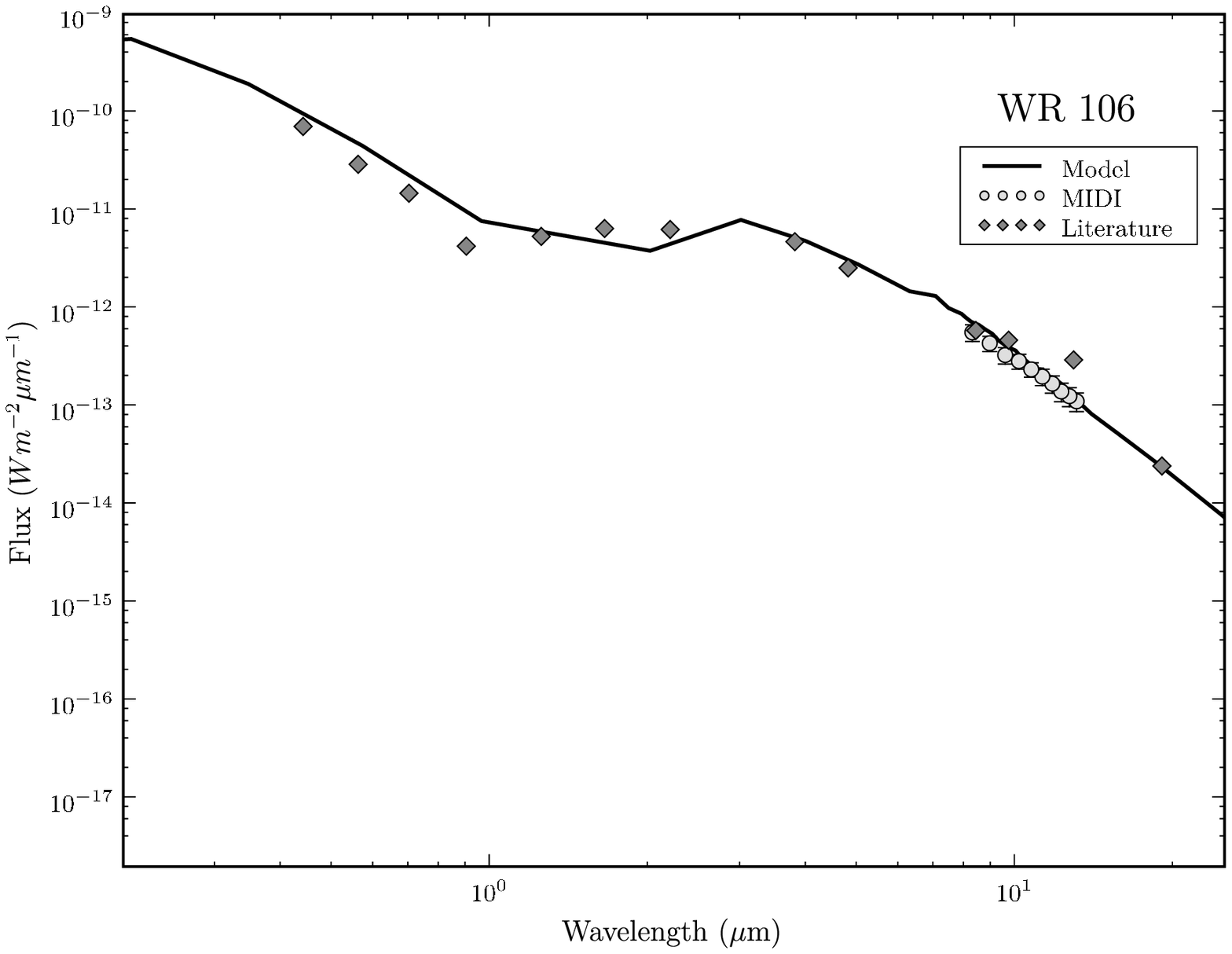}
\caption{SED of WR 106. The solid line is our best model, the grey circles are 
the measurements from MIDI, the squares are measurements from the literature (see text
for sources). \label{fig9}}
\end{figure}

\clearpage

\begin{figure}
\plotone{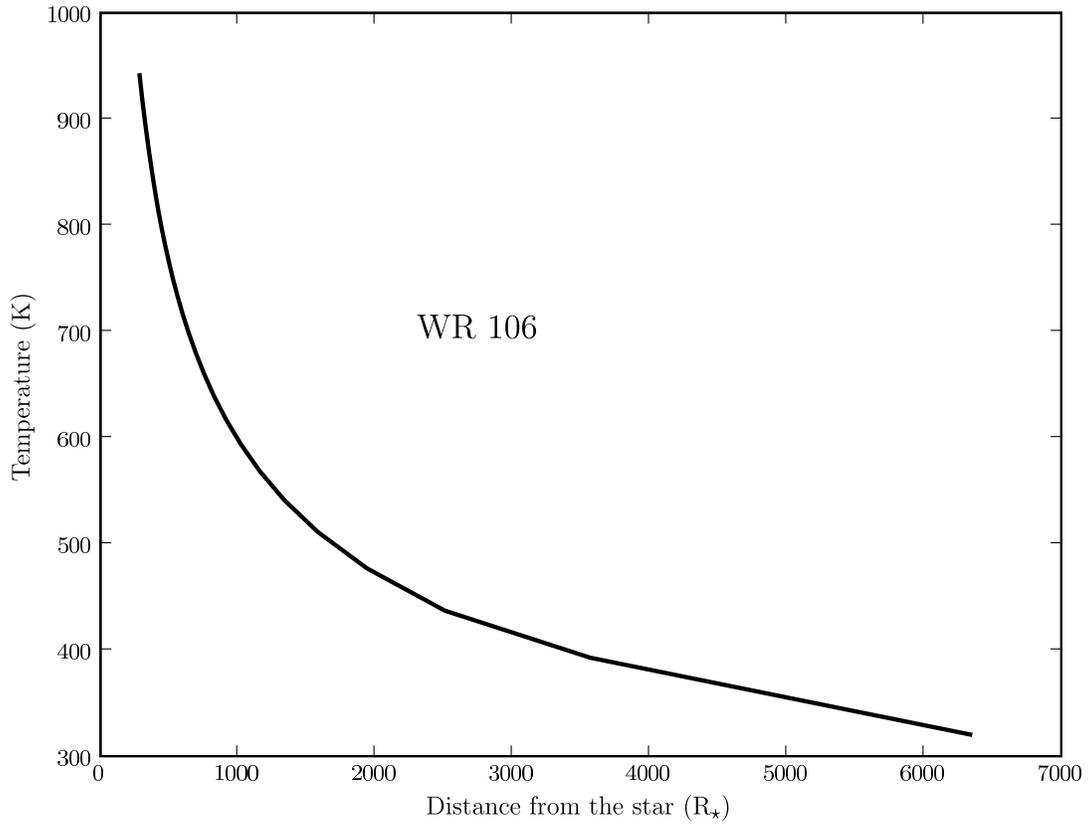}
\caption{Dust temperature profile for the WR 106 best model.\label{fig10}}
\end{figure}


\clearpage

\begin{figure}
\plotone{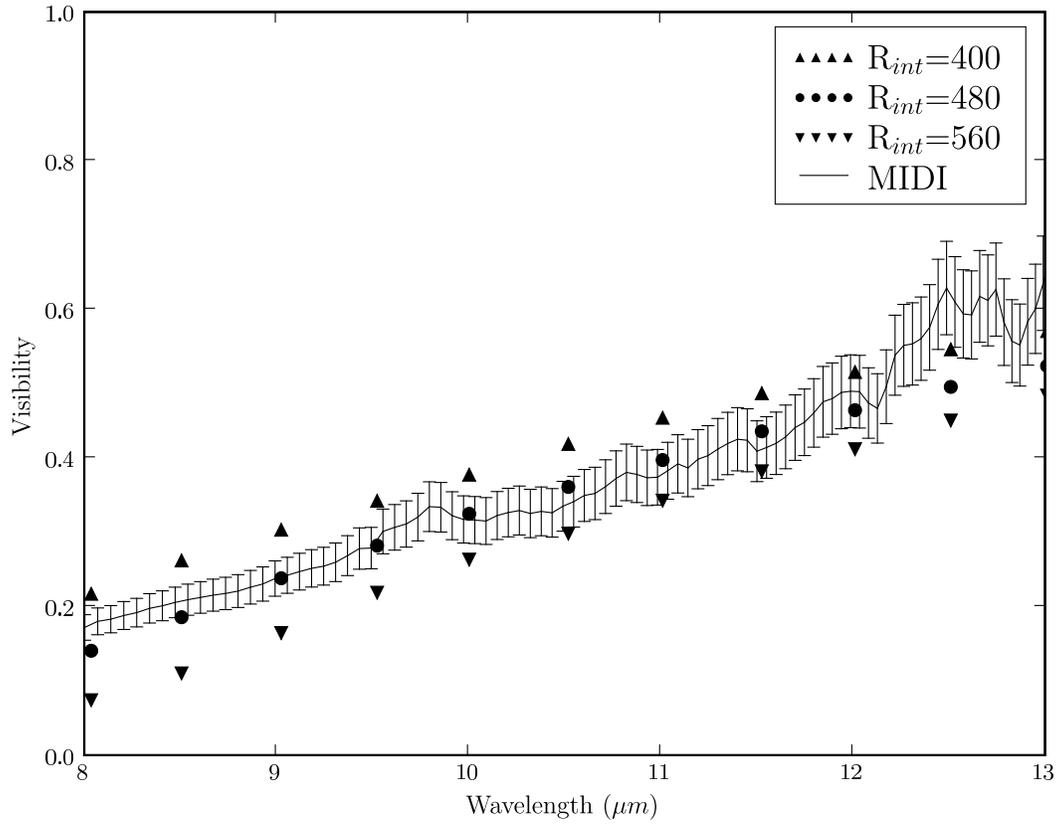}
\caption{Visibility versus wavelength for WR 95. The line (with error bars) is the MIDI measurement, the filled circles are the visibility points 
from the best model. For comparison we show visibilities for models with different inner radii (R$_{int}$) for the shell.\label{fig11}}
\end{figure}

\clearpage

\begin{figure}
\plotone{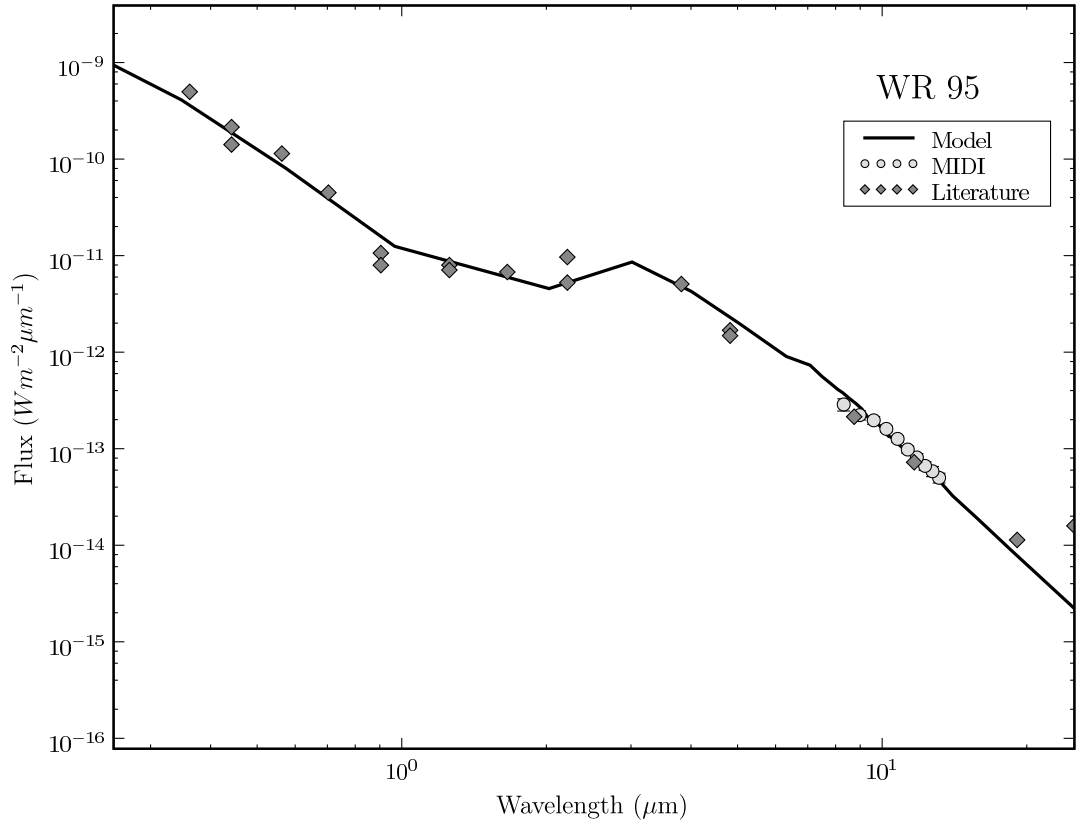}
\caption{SED for WR 95: The solid line is our best model, the grey circles the measurements from MIDI (de-reddened as in text), 
the diamonds are measurements from the literature (see text for sources)\label{fig12}}
\end{figure}

\clearpage

\begin{figure}
\plotone{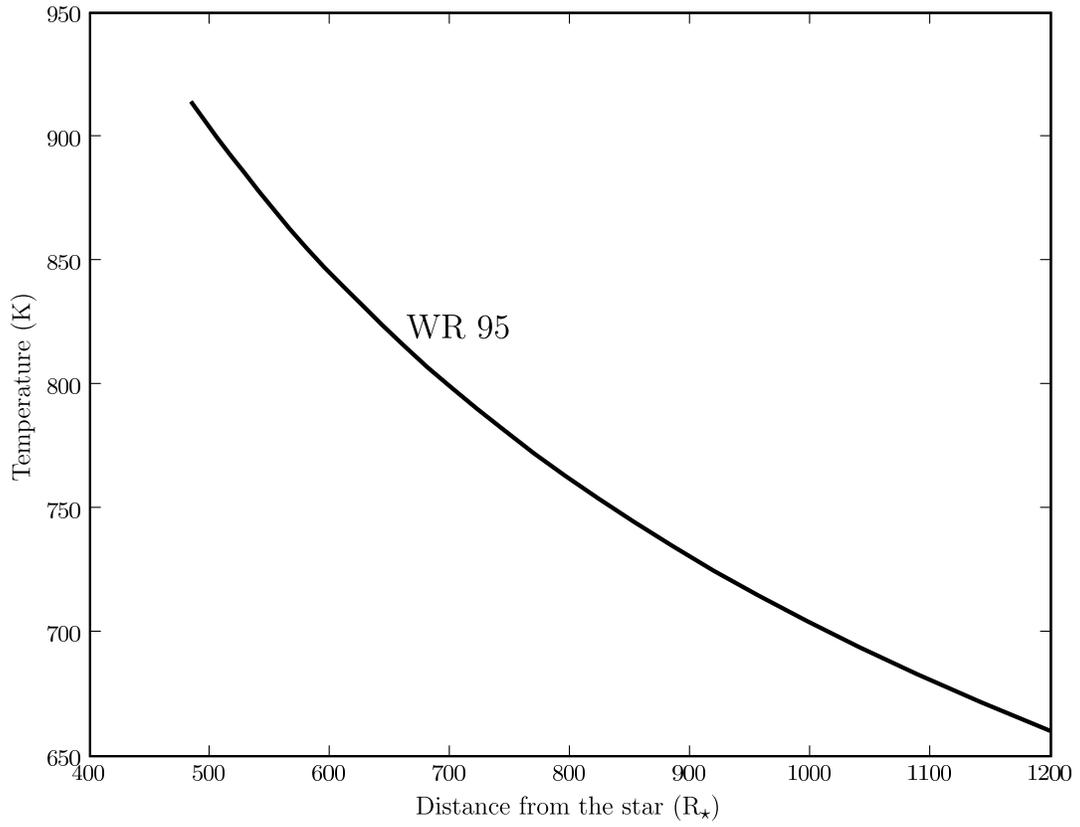}
\caption{Dust temperature profile of our WR 95 model.\label{fig13}}
\end{figure}

\end{document}